\documentclass[aps,prd,floatfix,notitlepage,noshowpacs,preprintnumbers,nofootinbib,superscriptaddress,twocolumn]{revtex4-2}

\usepackage[utf8]{inputenc}

\usepackage{tablefootnote}
\usepackage[x11names,dvipsnames]{xcolor}
\usepackage{listings}
\usepackage{siunitx}
\usepackage{graphicx}
\usepackage{dcolumn}
\usepackage{bm}
\usepackage{amsmath,amssymb}
\usepackage[version=4]{mhchem}
\usepackage{placeins}
\usepackage[caption=false]{subfig}
\usepackage{float}

\makeatletter
\def\switch@array{}
\makeatother

\newcommand{\CCCoh}{\nu_\mu\text{CC-Coh}\pi}

\newcommand{\uCCCoh}{\nu_\mu\text{CC-Coh}\pi}

\begin{document}

\preprint{FERMILAB-PUB-24-0584-PPD}

\title{Neutrino-induced charged-current coherent pion production for constraining the muon neutrino flux at DUNE}

\author{Mun Jung Jung}
 \email{munjung@uchicago.edu }
\affiliation{%
 University of Chicago, Chicago, Illinois 60637, USA
}%
\author{Vishvas Pandey}%
\affiliation{%
  Fermi National Accelerator Laboratory, Batavia, Illinois 60510, USA
}%
\author{Gray Putnam}%
 \email{gputnam@fnal.gov}
\affiliation{%
  Fermi National Accelerator Laboratory, Batavia, Illinois 60510, USA
}%
\author{David W. Schmitz}
\affiliation{%
 University of Chicago, Chicago, Illinois 60637, USA
}%


\begin{abstract}
We study neutrino induced charge current coherent pion production ($\CCCoh$) as a tool for constraining the neutrino flux at the Deep Underground Neutrino Experiment (DUNE). The neutrino energy and flavor in the process can be directly reconstructed from the outgoing particles, making it especially useful to specifically constrain the muon neutrino component of the total flux. The cross section of this process can be obtained using the Adler relation with the $\pi$-Ar elastic scattering cross section, taken either from external data or, as we explore, from a simultaneous measurement in the DUNE near detector. We develop a procedure that leverages $\CCCoh$ events to fit for the neutrino flux while simultaneously accounting for relevant effects in the cross section. We project that this method has the statistical power to constrain the uncertainty on the normalization of the flux at its peak to a few percent. This study demonstrates the potential utility of a $\CCCoh$ flux constraint, though further work will be needed to determine the range of validity and precision of the Adler relation upon which it relies, as well as to measure the $\pi$-Ar elastic scattering cross section to the requisite precision. We discuss the experimental and phenomenological developments necessary to unlock the $\CCCoh$ process as a ``standard candle'' for neutrino experiments. 

\end{abstract}

\maketitle

\section{Introduction}
The upcoming Deep Underground Neutrino Experiment (DUNE) will measure charge-parity (CP) violation in the neutrino sector by observing the oscillation of muon and electron (anti-)neutrinos with an energy $\sim 0.5$-\SI{5}{GeV} over a baseline of \SI{1300}{km}~\cite{DUNETDR}. These measurements are sensitive to the modeling of the neutrino flux, as well as neutrino-argon cross sections, both of which have significant uncertainties. The use of a near and far detector, as well as both a predominantly neutrino and predominantly anti-neutrino beam, partially control these uncertainties. Still, methods to reduce these uncertainties will enhance the power of DUNE to measure CP violation. 

The neutrino flux can be constrained in-situ by measuring the rate of a process with a well understood cross-section. This has been explored in the context of neutrino-electron elastic scattering; the technique has been demonstrated in the MINERvA experiment~\cite{MINERvAConstraint1, MINERvAConstraint2} and its constraining power has been projected for DUNE~\cite{DUNEConstraint}. In this paper, we explore the possibility of a neutrino flux constraint with the neutrino induced charged current coherent pion production process ($\CCCoh$). Such a constraint would be powerful on its own due to the high rate of neutrino-nucleus interactions in the DUNE near detector (ND). There will be about 270,000 $\CCCoh$ interactions per year in a \SI{68}{\tonne} fiducial volume in DUNE ND, compared to 7,500 neutrino-electron scattering events. A $\CCCoh$-based constraint would also complement a neutrino-electron scattering measurement by constraining different components of the neutrino flux. In particular, the flavor of the neutrino can be reconstructed in $\CCCoh$ interactions, unlike in the neutrino-electron scattering case. Furthermore, neutrino-electron interactions in the DUNE near detector can be used in other measurements such as the weak mixing angle~\cite{WeakMixing} and searches for hidden sector bosons~\cite{DarkZ}. A flux constraint with a different interaction channel would improve the reach of those measurements, and could also be used to disentangle flux modeling from any new physics signal.

Neutrino oscillation experiments have previously attempted other methods of using neutrino-nucleus interactions to constrain the flux such as inclusive charged current interactions~\cite{SciBooNEConstraint} and measuring interactions at low inelasticity (the ``low-$\nu$'' method)~\cite{Lownu1, LownuNuTeV, LownuNomad, LownuMINOS, LownuMINERvA1, LownuMINERvA2}. It has also been proposed to leverage anti-neutrino interactions on hydrogen as a constraint~\cite{HConstraint1, HConstraint2}. Such flux constraints must pass three requirements~\cite{LownuBad}.
\begin{enumerate}
    \item That the cross section of the process can be obtained in a model-independent way.
    \item That the neutrino energy can be reconstructed in a model-independent way from measurements of the final state particles.
    \item That a sample of events in the interaction can be obtained with high purity and with small uncertainties on the detector performance.
\end{enumerate}

Cross section measurements of the $\CCCoh$ process demonstrate the feasibility of requirement 3 (e.g., Ref.~\cite{CohPixsecCHARM-II, CohPixsecMINERvA1, CohPixsecMINERvA2, CohPixsecMINERvA3, CohPixsecT2K}). The DUNE ND will be a pixelated liquid argon time projection chamber (LArTPC)~\cite{DUNETDR}, which projects to have a particularly strong ability to isolate this process. Background incoherent neutrino interactions can be rejected by vetoing on additional particles to the $\mu\pi^\pm$ pair, which in LArTPCs can be identified down to low energies:  $\sim$\SI{15}{MeV} for protons~\cite{ICARUSDiMuon} and $\sim$\SI{50}{MeV} for muons and pions~\cite{MicroBooNEMarcoXSec}. Identifying the final state $\gamma$-rays produced in incoherent neutrino interactions would provide a further veto~\cite{ArgoNeuTLowE, Blips}.

Requirement 2 is passed because the interaction is coherent and therefore the argon nucleus stays in its ground state. A tiny fraction of the neutrino energy is taken up by the nucleus recoil and the rest is measurable from the outgoing muon and pion. 

This leaves requirement 1, that the process has a known cross section. This requirement can be addressed using the Adler relation, derived from the partially conserved axial-vector current (PCAC) theorem~\cite{Adler1, Adler2}
\begin{equation}
    \frac{d\sigma^{CC}}{dQ^2dy d|t|}\biggr\rvert_{Q^2 = 0, m_l = 0}^{\rm PCAC} = \frac{G_F^2}{2\pi^2}  f_\pi^2 \cos^2\theta_C \frac{(1-y)}{y} \frac{d\sigma_{el}^{\pi A}}{d|t|},
\end{equation}
where $|t| = |(q-p_{\pi})^2| = |(p_{\nu}-p_{\mu}-p_{\pi})^2|$, $y = (E_\nu - E_\mu)/E_\nu \approx E_\pi / E_\nu$, $Q^2 \simeq ({p_\nu} - {p_\mu})^2$, 
$G_F$ is the Fermi coupling constant, $\theta_C$ is the Cabbibo angle, $f_\pi$ is the pion decay constant, and $\frac{d\sigma_{el}^{\pi A}}{d|t|}$ is the cross section for pion-nucleus elastic scattering. This relation relies on the elastic pion-nucleus cross section, which has not been directly measured in argon. Previous attempts to apply the Adler relation to $\CCCoh$ scattering on argon have built up a pion-nucleus model from pion-nucleon scattering~\cite{ReinSehgal}, or used data on lighter nuclei such as carbon~\cite{BergerSeghal, Paschos1}. In order for the $\CCCoh$ cross section to be understood precisely, a direct measurement of $\pi$-Ar elastic scattering is necessary. While challenging, this measurement could potentially be performed in existing experiments such as ProtoDUNE-SP, ProtoDUNE-HD~\cite{ProtoDUNE1, ProtoDUNE2}, or LArIAT~\cite{LArIAT}. LArIAT has previously measured inclusive $\pi$-Ar scattering~\cite{LArIATxsec}. 

In this paper, we furthermore propose the use of a different experiment to measure the elastic cross section: \textit{DUNE}. The DUNE near detector (ND) can measure the elastic $\pi$-Ar  cross section with the pions produced in neutrino interactions. The wide range of energies of pions produced in neutrino interactions allow the full range of kinematic space relevant for $\CCCoh$ scattering to be measured. The full kinematics of the elastic$\pi$-Ar  interaction can be reconstructed event-by-event from the pion deflection angle and its energy deposited in the detector (this is uniquely true for the elastic interaction, it would not work for any inelastic process without a priori knowledge of the pion energy). There will be millions of $\pi$-Ar elastic scatters in DUNE ND, so in principle the detector can measure the cross section precisely (up to any systematic uncertainties).

After the $\pi$-Ar elastic cross section is obtained, the Adler relation reduces to the ratio
\begin{equation}\label{eqn:cohpiratio}
    \frac{\frac{d\sigma^{CC}}{dQ^2dy d|t|}\bigr\rvert_{Q^2 = 0, m_l = 0}^{\rm PCAC}}{\frac{d\sigma_{el}^{\pi A}}{d|t|}} = \frac{G_F^2}{2\pi^2}  f_\pi^2 \cos^2\theta_C \frac{(1-y)}{y},
\end{equation}
where now the right hand side consists of fundamental constants that are measured very precisely by other experiments: $G_F^2$ from muon decay~\cite{PDG, MuonLifetime}, and $f_\pi^2\cos^2\theta_C$ from pion decay~\cite{PionDecayConstant}. This ratio can be used to constrain the neutrino flux. If both the numerator and denominator can be obtained in DUNE ND, then systematic uncertainties may be reduced in the fraction, such as any uncertainty in the detector performance or pion energy scale.

Care must be taken to apply the Adler relation in a region of phase space where higher order effects are not too large~\cite{PCACBad, Paschos1}. For this work, we apply estimates of the valid regions of phase space from previous studies of the $\CCCoh$ process. Future studies that can more precisely quantify the kinematic regime where the Adler relation does (or does not) apply and compute corrections or uncertainties for any departure(s) will be necessary to apply this technique in practice at DUNE. Measurements of the process at ongoing and upcoming neutrino experiments such as SBN~\cite{SBN1, SBN2} can provide complimentary experimental tests of the Adler relation independent of any application at DUNE. In addition to its potential use as a flux constraint, measurements of the $\CCCoh$ process shed light on the fundamental nature of the axial current in neutrino interactions, as expressed by the PCAC theorem~\cite{Adler1, Adler2}. In addition, this process has similar kinematics to particle decays to $\mu\mu$ and $\mu\pi^\pm$ final states, and so can serve as an important background to beyond standard model physics searches~\cite{ICARUSDiMuon}.

In this work, we develop a methodology to use the $\CCCoh$ process as an in-situ constraint of the muon neutrino flux in the DUNE near detector. We demonstrate through bias tests that this procedure is resilient against possible systematic effects in the cross section. We estimate the performance of DUNE ND by applying results from previous LArTPC detectors, as well as simple simulations of the near detector. With this estimation, we project the power of the flux constraint for DUNE. 

The paper is organized as follows. In Sec.~\ref{sec:CCCohPixsec}, we detail how we model the $\CCCoh$ cross section and what phase space restrictions we make. In Sec.~\ref{sec:method}, we describe the flux constraint methodology. In Sec.~\ref{sec:result}, we show the result of the constraint, including bias tests. Finally, Sec.~\ref{sec:conclusion} concludes the paper.

\section{Neutrino Induced Charged Current Coherent Pion Production}
\label{sec:CCCohPixsec}
\begin{figure}
    \centering
    \includegraphics[width=0.80\linewidth]{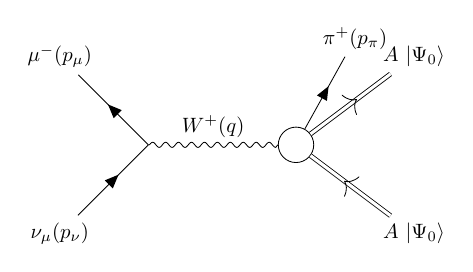}
    \caption{Diagrammatic representation of the muon neutrino induced CC coherent pion production considered in this work, where a single $W^+$ boson is exchanged between the neutrino and the target nucleus. The nucleus remains in its ground state, while a muon and a charged pion are produced in the forward direction.}
    \label{fig:feynman_diagram}
\end{figure}

Coherent pion production refers to the process in which a(n) (anti-)neutrino scatters off a nucleus, producing a forward pion while leaving the nucleus in its ground state. The process is termed ``coherent" because the overall scattering amplitude results from the constructive interference of the scattering amplitudes of the incident wave on individual nucleons within the target nucleus, leading to an enhanced cross section. Since the incident wave is approximately the same on all nucleons, the target nucleus recoils as a whole without breaking up and with very little recoil energy, and no quantum numbers (charge, spin or isospin) are transferred to the target nucleus. Such processes can occur through both charged current (CC) and neutral current (NC) induced reactions. In this study, we focus on muon neutrino induced charged current coherent pion production on argon
\begin{equation}
    \nu_\mu ({p_\nu}) + \rm{A} \rightarrow \mu^- ({p_\mu})  + \pi^+ ({p_\pi})  + \rm{A} .
\end{equation} 
The process is schematically shown in Fig.~\ref{fig:feynman_diagram}. A muon neutrino with four-momentum ${p_\nu}$ scatters off an argon nucleus, producing a forward-going muon with four-momentum ${p_\mu}$ and a charged pion with four-momentum ${p_\pi}$, while the argon nucleus remains in its ground state. 

In the coherent pion production process, the four momentum transferred from the leptonic current to the nucleus, $q = p_{\nu} - p_{\mu}$, is very small. We treat the nucleus as infinitely heavy so that all the energy loss at the neutrino vertex, $\nu$, is transferred to the outgoing pion, such that $\nu = E_\nu - E_\mu$, where $E_\nu$ and $E_\mu$ are the energies of the muon neutrino and the muon in the lab frame. Since the nucleus spin is not flipped during coherent scattering and its recoil is minimal, we have $\nu \simeq E_\pi$, where $E_\pi$ is the pion energy in the lab frame. The process is characterized by a small four-momentum transfer to the nucleus and an exponential decrease in the cross section with $|t|$, the four-momentum transfer between the incoming virtual boson and the outgoing pion. 

Experimentally, the NC and CC coherent pion production processes have been observed in various nuclei across medium and high energy ranges. The first such measurement was reported by the Aachen-Padova collaboration in 1983~\cite{Faissner:1983ng}, during their study of isolated $\pi^0$s produced in $\nu_\mu$ and $\bar{\nu}_\mu$ induced processes. This was followed by the Aachen-Gargamelle group, who isolated coherent NC $\pi^0$ events in the Gargamelle heavy freon exposure~\cite{Isiksal:1984vh}. Subsequent neutrino experiments, such as CHARM~\cite{CHARM:1985bva, CHARM-II:1993xmz} and SKAT~\cite{SKAT:1985uch, Nahnhauer:1986xh}, observed NC-induced coherent pions over a wide range of neutrino energies using different nuclear targets. More recently, several accelerator-based neutrino experiments, including K2K~\cite{K2K:2005uiu}, SciBooNE~\cite{SciBooNE:2008bzb}, MiniBooNE~\cite{MiniBooNE:2008mmr}, NOMAD~\cite{NOMAD:2009idt}, ArgoNeuT~\cite{ArgoNeuT:2014uwh}, T2K~\cite{T2K:2016soz}, MINERvA~\cite{MINERvA:2014ani, MINERvA:2022esg}, MINOS~\cite{MINOSxsec}, and NOvA~\cite{NOvA:2019bdw}, have either established limits on the CC-induced coherent pion production cross section or provided direct measurements.

Theoretically, the invariant matrix element of the charged current process shown in Fig.~\ref{fig:feynman_diagram} can be written as
\begin{equation}\label{matrix_element}
    \mathcal{M} = - \frac{G_F \,\rm{cos}\theta_C}{\sqrt{2}} j_\mu \langle \pi^+ A| J^{\mu} | A\rangle,
\end{equation}
where $j_\mu$ is ($\nu_\mu \rightarrow \mu$)-matrix element of the leptonic current expressed as
\begin{equation}
    j_\mu = \bar{u} (p_\mu) \gamma_\mu (1-\gamma_5) u (p_\nu).
\end{equation}
The hadronic matrix element can be derived using Adler’s PCAC theorem~\cite{Adler1, Adler2, Adler3}. As a result of PCAC, the longitudinal component of the axial-vector current couples to the pion field with a strength proportional to the pion decay $(\pi \rightarrow \mu \nu_{\mu})$ constant, $f_\pi (\approx 0.93 m_\pi)$. Under the PCAC assumption, Adler's relation provides a relationship between the hadronic matrix element for neutrino-induced pion production and the pion-nucleus elastic scattering amplitude at $Q^2 = -q^2 = -(p_\nu - p_\mu)^2 = 0$. In this context, the vector current contribution is not only suppressed by a factor of $1/\nu$ but is also forbidden by quantum number selection rules in coherent processes. The first calculation based on this approach was carried out by Rein and Sehgal~\cite{Rein:1980wg}, with subsequent work by the same group and by others over the years~\cite{ReinSehgal, Rein:2006di, Berger:2007rq, BergerSeghal, Belkov:1986hn, Paschos1, Paschos2}. In this framework, the differential cross section for CC single pion production is given by
\begin{equation}\label{eq:cs_adler_1}
    \frac{d\sigma^{CC}}{dQ^2dy d|t|}\biggr\rvert_{Q^2 = 0, m_l = 0}^{\rm PCAC} = \frac{G_F^2}{2\pi^2}  f_\pi^2 \cos^2\theta_C \frac{(1-y)}{y} \frac{d\sigma_{el}^{\pi A}}{d|t|}.
\end{equation}

Being proportional to the elastic pion-nucleus cross section, $\frac{d\sigma_{el}^{\pi A}}{d|t|}$, the differential cross section in Eq.~\ref{eq:cs_adler_1} shows a sharp exponential decrease with $|t|$. 

The expression above was initially derived assuming massless leptons, even in the case of the CC process. In subsequent work, Rein and Sehgal~\cite{Rein:2006di} and Berger and Sehgal~\cite{Berger:2007rq} incorporated lepton mass effects stemming from the pion-pole term in the hadronic axial-vector current. Including this correction is essential for the CC process, as it leads to a suppression of the cross section at forward muon scattering angles, addressing the deficit observed in experimental data (e.g., K2K data) at $Q^2 \lessapprox 0.1, \text{GeV}^2$. When the muon mass, $m_\mu$, is not neglected, the reaction receives a contribution from the exchange of a charged pion between the lepton vertex and the hadron vertex. The coupling at the lepton vertex and the amplitude contains the characteristic pion propagator, $(Q^2 + m_\pi^2)^{-1}$. This pseudoscalar amplitude interferes with the remaining amplitude, which is free of the pion singularity; Rein and Sehgal in Ref.~\cite{Rein:2006di} called pole-free contribution the ``axial'' amplitude. These two amplitudes interfere destructively. The correction factor, known as Adler's screening effect, is expressed as:
\begin{equation}
\label{eq:zeta}
\begin{split}
\xi_{\rm Adler} = & \left( 1 - \frac{Q^2_{\rm{min}}}{2(Q^2 + m_\pi^2)} \right)^2 \\
& + \frac{y}{4} Q^2_{\rm{min}} \left(\frac{Q^2 - Q^2_{\rm{min}}}{(Q^2 + m_\pi^2)^2} \right) 
\end{split}
\end{equation}
where $Q^2_{\rm{min}} = m_\mu^2 y /(1-y)$ and the range of the variable $Q^2$ is $Q^2_{\rm{min}} \le Q^2 \le 2\,M\,E_\nu y_{\rm{max}}$ where $y$ lies between $y_{\rm{min}} = m_\pi / E_\nu$ and $y_{\rm{max}} = 1 - m_\mu / E_\nu$. This destructive interference is evident in the first term of the correction factor in Eq.~\ref{eq:zeta}. The two terms within the parentheses represent the axial and pseudoscalar amplitudes, with the negative sign indicating destructive interference, resulting in suppression at low-$Q^2$. This effect occurs exclusively in charged current scattering, where the muon mass plays a significant role. The fact that the muon mass is comparable to the pion mass in the pion propagator is crucial. However, the effect diminishes as neutrino energy increases. The neutral current channels remain unaffected. The first term in Eq.~\ref{eq:zeta} corresponds to outgoing muons with negative helicity (helicity nonflip) while the second term represents the helicity flip contribution, which vanishes at a scattering angle of $0^{\circ}$. When the lepton mass is neglected, the additional term multiplied by the lepton current contributes zero and has no effect. However, if  $m_\mu \neq 0$, the pion-pole term does contribute. Thus the lepton mass corrected PCAC formula, applicable for small-angle scattering, is given by
\begin{equation}
\label{eq:cs_adler_2}
\begin{split}
    \frac{d\sigma^{CC}}{dQ^2dy d|t|}\biggr\rvert_{Q^2 \approx 0, m_l \neq 0}^{\rm PCAC}
    = {} &  \frac{G_F^2}{2\pi^2}  f_\pi^2 \cos^2\theta_C \frac{(1-y)}{y} \frac{d\sigma_{el}^{\pi A}}{d|t|} \\
    & \times \xi_{\rm Adler}
\end{split}
\end{equation}
The effect of lepton mass correction has been studied in previous literatures. We quantitatively compared the lepton mass correction terms from the models of Kartavtsev–Paschos~\cite{Paschos1} and Berger–Sehgal~\cite{BergerSeghal} as a function of $Q^2$ and $t$, and found differences at the level of a few percent maximum in the relevant kinematic region. This confirms that the treatment follwed by this study remains consistent in magnitude and trend with alternative approaches, though future studies may be valuable.

While the process remains coherent for small, non-zero values of $Q^2$, another problem appears in how to extrapolate the above relation to a finite
$Q^2$ value. The Rein-Sehgal scheme~\cite{ReinSehgal} addresses this by introducing a propagator term through an axial form factor, $\mathcal{F_A} (Q^2)$, which is incorporated into Eq.~\ref{eq:zeta} as
\begin{equation}
\label{eq:cs_adler_3}
\begin{split}
    \xi_{\rm Adler} = & \left(\mathcal{F_A} (Q^2)  - \frac{Q^2_{\rm{min}}}{2(Q^2 + m_\pi^2)} \right)^2 \\
    & + \frac{y}{4} Q^2_{\rm{min}} \left(\frac{Q^2 - Q^2_{\rm{min}}}{(Q^2 + m_\pi^2)^2} \right) 
\end{split}
\end{equation}
resulting a PCAC based cross section that is extended to $m_l \neq 0$ and $Q^2 \neq 0$ and stays as a valid approximation for low $Q^2$ values. Note that the hadronic matrix element in Eq.~\ref{matrix_element} is reduced to a combination of $\frac{d\sigma_{el}^{\pi A}}{d|t|}$ and $\mathcal{F_A} (Q^2)$. 

Finally, while Eq.~\ref{eq:cs_adler_2} provides a convenient relation between the coherent pion production amplitude and the pion–nucleus elastic scattering cross section, it is formally valid only for an on-shell exchanged pion. In the actual weak interaction process, however, the axial current couples to an off-shell pion. Using on-shell pion–nucleus cross-section data to model coherent production is therefore an approximation that neglects off-shell effects. Nevertheless, in the absence of a fully first-principles description of the nuclear response for argon, employing pion–argon elastic scattering data remains a practical and widely adopted approach. In the CC channel, the presence of the outgoing muon imposes a non-zero minimum $Q^2$, which suppresses the kinematic region where the pion is far off-shell. This makes the on-shell approximation somewhat better justified for CC coherent pion production at moderate neutrino energies, with residual off-shell effects effectively absorbed into the phenomenological factor $\xi_{\rm Adler}$.

In addition to the models based on Adler's relation discussed above~\cite{Rein:1980wg, ReinSehgal, Rein:2006di, Berger:2007rq, BergerSeghal, Belkov:1986hn, Paschos1, Paschos2}, there is a second category of approaches that rely on more microscopic models for pion production~\cite{Kelkar:1996iv, Singh:2006br, Alvarez-Ruso:2007rcg, Alvarez-Ruso:2007kwp, Amaro:2008hd, Leitner:2009ph, Hernandez:2009vm, Zhang:2012xi, Nakamura:2009iq}. These models are based on the single-nucleon process, dominated by $\Delta$ production within the nucleus. The total cross section is obtained by coherently summing the contributions of the pion production amplitudes from all nucleons in the nucleus. These approaches account for the nuclear medium modification of $\Delta$ properties in the nucleus, as well as the final state interactions (FSI) of the outgoing pion with the nuclear target. In principle, PCAC-based models should emerge as an approximation to these more microscopically motivated nuclear structure models, in particular for higher neutrino energies and low $Q^2$ processes~\cite{Leitner:2009ph}. 

In this work, we utilize the Berger-Sehgal model based on Adler's relation, Eq.~\ref{eq:cs_adler_2}, due to its simplicity.
The PCAC prescription for the $\CCCoh$ cross section is reliable in cases where the transverse component of the axial current is minimized. We also furthermore limit the phase space used in accordance with previous theoretical studies of the interactions which require that $\nu > 3\sqrt{Q^2}$ and $Q^2 \lesssim$ \SI{0.2}{GeV\squared}~\cite{Paschos1,Paschos2}. Within these constraints, we assume that Eq.~\ref{eq:cs_adler_2} holds perfectly for this study. This equation relies on two inputs: the form factor $\mathcal{F_A} (Q^2)$ and the pion-nucleus elastic scattering cross section $\frac{d\sigma_{el}^{\pi A}}{d|t|}$. These are detailed below in Sections~\ref{sec:FF} and~\ref{sec:pixsec}, respectively.

Finally, we note that future work that can assess in a precise way whether PCAC-breaking effects in the forward scattering limit (e.g., on-shell pion approximation, transverse currents) on $^{40}$Ar spoil the Adler relation, or otherwise necessitate additional systematic uncertainty, will be integral in order to leverage this process as a flux constraint in DUNE.

\subsection{Form Factor}
\label{sec:FF}

A form factor $\mathcal{F_A} (Q^2)$ extrapolates the Adler relation to finite $Q^2$. From the assumption that the axial current is dominated by a heavy meson (e.g., the $a_1$~\cite{PDG}), Berger-Sehgal used a dipole form of the form factor as in~\cite{BergerSeghal}

\begin{equation*}
    \mathcal{F_A} (Q^2) =  \frac{\mathcal{F_A} (0)}{(1+Q^2/m_A^2)^2}
\end{equation*}
with $m_A \approx 1.0$ GeV as the axial mass. The Belkov-Kopehovlch approach~\cite{Belkov:1986hn}, which is based on a dispersion generalization of the Adler relation, employs the Glauber model to introduce $Q^2$ dependence and account for the non-resonant background. The primary contribution to the non-resonant background in this reaction is associated with ($\rho \pi$) pair production~\cite{Belkov:1986hn}. Assuming that non-resonant ($\rho \pi$)-systems dominate the axial-vector current, a cut term is obtained instead of the dipole factor. In this scenario, the dependence of the spectral function of the dispersion relation on $\Lambda^2$ (the squared effective mass of the ($\rho \pi$)-system) is defined by the
factor $(\Lambda^2 - m_\pi^2)^{-1}$. Thus the cut contribution to the axial form factor, normalized to unity at $Q^2 = 0$, is expressed as follows:
\begin{eqnarray*}
   \mathcal{F}_{\rm cut}(Q^2) & = & \mathcal{F_A}(0) \int_{(m_\rho + m_\pi)^2}^{\infty} d\Lambda^2 \frac{(m_\rho + m_\pi)^2}{(Q^2 + \Lambda^2) (\Lambda^2 - m_\pi^2)} \nonumber \\
                &  = & \mathcal{F_A}(0) \frac{(m_\rho + m_\pi)^2}{Q^2 + m_\pi^2}\ln\left[1 + \frac{Q^2 + m_\pi^2}{(m_\rho + m_\pi)^2}\right]\, .
\end{eqnarray*}
For small values of $Q^2$, $\mathcal{F_A}(Q^2)$ and $\mathcal{F}_{\rm cut}(Q^2)$ are approximately equal while the cross-section normalization at $Q^2 = 0$ is fixed by Adler's relation. Consequently, as long as $Q^2$ is kept to smaller values, the form factor contribution from $\rho \pi$ production in the intermediate state can be neglected. 

However, for a precise model of $\CCCoh$ production it is not clear that such simple prescriptions for the form factor are adequate. As discussed above, at finite $Q^2$ other currents perturb the cross section, and the form factor implicitly folds in these additional contributions. Other neutrino interaction models have addressed similar inadequacies in the dipole form factor with the $z$-expansion approach~\cite{ZExp2, ZExp3}. By expressing the form factor in a general way as a convergent power expansion of $z$,
\begin{equation}
    \label{eq:zexpFF}
    \mathcal{F_A} (\vec{a}, Q^2) = \sum_{k=0}^{k_{max}} \vec{a}_k z(Q^2)^k,
\end{equation}
where $\vec{a}_k$ are dimensionless coefficients, the form factor can be represented in a model-independent way \cite{ZExp1}. The variable $\textit{z}$ is a function of $Q^2$,
\begin{equation}
    z(Q^2, t_{cut}, t_0) = \frac{\sqrt{t_{cut} + Q^2} - \sqrt{t_{cut} - t_0}}{\sqrt{t_{cut} + Q^2} + \sqrt{t_{cut} - t_0}},
\end{equation}
where $t_{cut}$ is the mass of of the lightest state that can be produced by the axial current, $k_{\rm max}$ is the total number of coefficients to be used for the expansion, and $t_0$ is a free parameter that is chosen to optimize the convergence of the expansion~\cite{ZExp2, ZExp3}. While the $z$-expansion model is not strictly necessary for the $\CCCoh$ cross section, it is adopted in this study as a flexible and familiar formalism to describe the extension of Adler's relation to small $Q^2$ values. 

\subsection{Pion-Argon Elastic Scattering Cross Section}
\label{sec:pixsec}
The pion-nucleus elastic cross section is the critical external input to the Adler relation to determine the strength of the $\CCCoh$ cross-section at $Q^2=0$. This cross section has never been measured on argon. (An inclusive cross section measurement, including elastic scatters where the deflection angle ($\theta_\pi$) is greater than $5^\circ$, has been performed by the LArIAT experiment~\cite{LArIATxsec}, and exclusive scattering measurements have been performed by the ProtoDUNE experiment~\cite{ProtoDUNExsec}.) Direct measurements, either by external detectors such as ProtoDUNE and LArIAT or, as we highlight, by DUNE itself, are needed to apply this process in a flux constraint. 

The pion-nucleus elastic cross section consists of an electromagnetic and QCD component that add at the amplitude level. Since \ce{^{40}Ar} is not an isoscalar, the QCD component is not the same between positive and negative pions. The sign of the pion also impacts the interference of the QCD and electromagnetic components of the cross section, which is important at small scattering angle. Measurements on similar nuclei such as iron indicate that the sign difference in the cross section is $\mathcal{O}(5\%)$  for $T_\pi >$~\SI{250}{MeV}~\cite{PiNXsecs}, well within the energy region of interest. Thus, if the $\pi$-Ar measurement is made in DUNE (which cannot measure the sign of the pion), correcting the cross section value by the relative fraction of the $\pi^+$ and $\pi^-$ flux would be necessary. If the uncertainty on the flux fraction can be kept moderately small ($\lesssim 20\%$), then the correction would not contribute a significant uncertainty.
External measurements of the cross section would not suffer from this uncertainty, and would also be useful to establish the precise magnitude of the sign difference on argon as a correction in a DUNE ND measurement.

In this work, we detail a flux constraint method that applies an entirely data-driven measurement of the $\pi$-Ar elastic cross section. However, leveraging a model for the process could be a useful augmentation to the technique. For example, such a model could reduce the uncertainty by constraining the possible kinematic dependence of the cross section, or interpolate to areas of phase space that cannot be directly constrained (e.g., small $\theta_\pi$). General purpose elastic scattering models such as those applied by GEANT4~\cite{GEANT4, G4DiffuseElastic} are capable of broadly describing the cross section across all nuclei. Measurements on \ce{^{12}C} have been modeled precisely through a framework that incorporates the partial wave expansion of the hadronic component of the cross section~\cite{C12PiEl1, C12PiEl2}. Such a model could be especially useful for the flux constraint since the diminishing of partial wave components at higher orders in principle allows high-angle scattering data to constrain lower angles.

\section{Methodology}
\label{sec:method}
Our proposed method obtains the flux constraint while simultaneously accounting for the systematic uncertainties on the $\CCCoh$ cross section model. The flux constraint is achieved by fitting for neutrino energy dependent scale factors to the 2D distribution of reconstructed neutrino energy and $Q^2$ of $\CCCoh$ events. The event selection for these events is outlined in Sec.~\ref{sec:eventsel}. The template fit approach used for this work, further detailed in Sec.~\ref{sec:flux_constraint}, resembles that of Ref.~\cite{DUNEConstraint}, where this method was used for the case of flux constraint with neutrino-electron elastic scattering events. 

We incorporate into the fit the two factors that determine the $\CCCoh$ cross section: the axial-vector form factor and the $\pi$-Ar elastic scattering cross section. The axial-vector form factor is modeled with a $z$-expansion formalism, and the coefficients are varied in the fit. The $\pi$-Ar elastic cross section is included in the fit in two separate ways to either emulate a DUNE ND (``DUNE ND only" scenario) or an external constraint (``external pion scattering data" scenario). For the DUNE ND case, we simultaneously perform a fit of the cross section to $\pi$-Ar elastic scattering events, binned by pion energy and scattering angle. For the external case, we include the magnitude of the $\pi$-Ar elastic cross section as a nuisance parameter with an assumed prior uncertainty. 

\subsection{Event Generation and Selection}
\label{sec:eventsel}
\subsubsection{$\CCCoh$ Event Selection}

\begin{figure}[]
    \centering
    \includegraphics[width=0.49\textwidth]{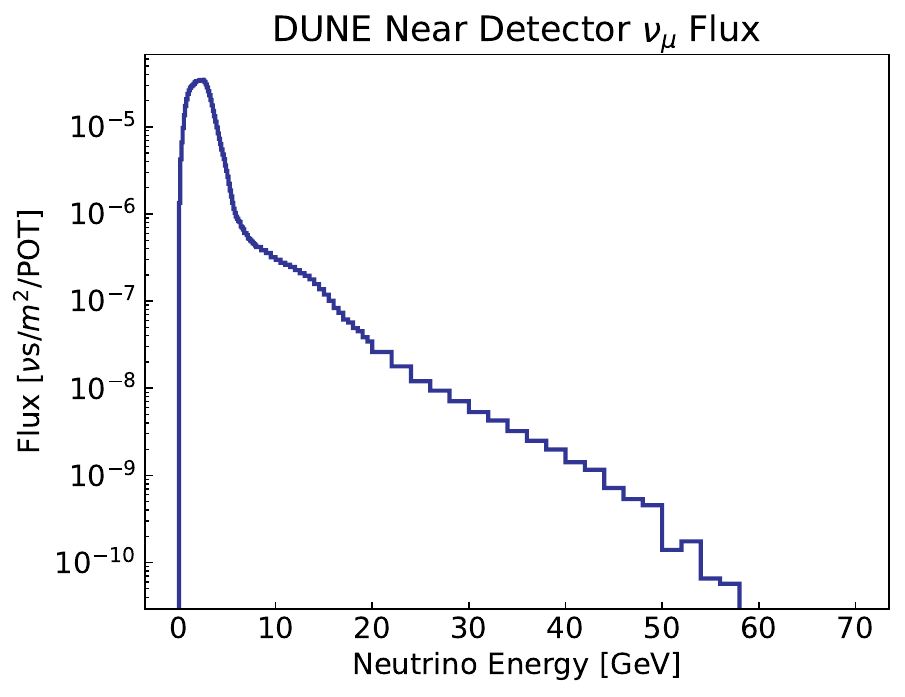}
    \caption{DUNE three-horn optimized flux for muon neutrino flavor in the forward horn current (FHC) mode.}
\label{fig:DUNE_numu_flux}
\end{figure}

We consider DUNE ND~\cite{DUNENDCDR} to be a liquid argon time projection chamber (LArTPC) with an active volume of \SI{7}{m} × \SI{3}{m} × \SI{5}{m}. The fiducial volume is defined with \SI{50}{cm} insets in the drift and vertical dimensions, \SI{25}{cm} in the front of the detector, and \SI{1.75}{m} in the back of the detector along the beam axis. The conservative cut of the fiducial volume at the back of the detector ensures that energy depositions from pions mostly stay contained in the detector. The neutrino flux used for this study is the DUNE three-horn optimized muon neutrino flux~\cite{DUNETDR}, shown in Fig.~\ref{fig:DUNE_numu_flux}. A total of 5 years of on-axis exposure with 1.1$\times 10^{21}$ protons on target (POT) per year is assumed to normalize the statistics.

A Monte Carlo simulation of neutrino interactions was generated using the GENIE neutrino event generator version v3.0.6 G18$\_$10a$\_$02$\_$11a~\cite{GENIE}. The GENIE event generator uses the Berger-Sehgal model for $\CCCoh$ events with pion kinetic energies lower than \SI{1}{GeV}~\cite{BergerSeghal}, and uses the Rein-Sehgal model for higher energies~\cite{ReinSehgal}. GENIE models the  form factor using the dipole parameterization with $m_A =\,$\SI{1}{GeV} in the default configuration used for this study. 

\begin{figure}[]
    \centering
    \includegraphics[width=0.49\textwidth]{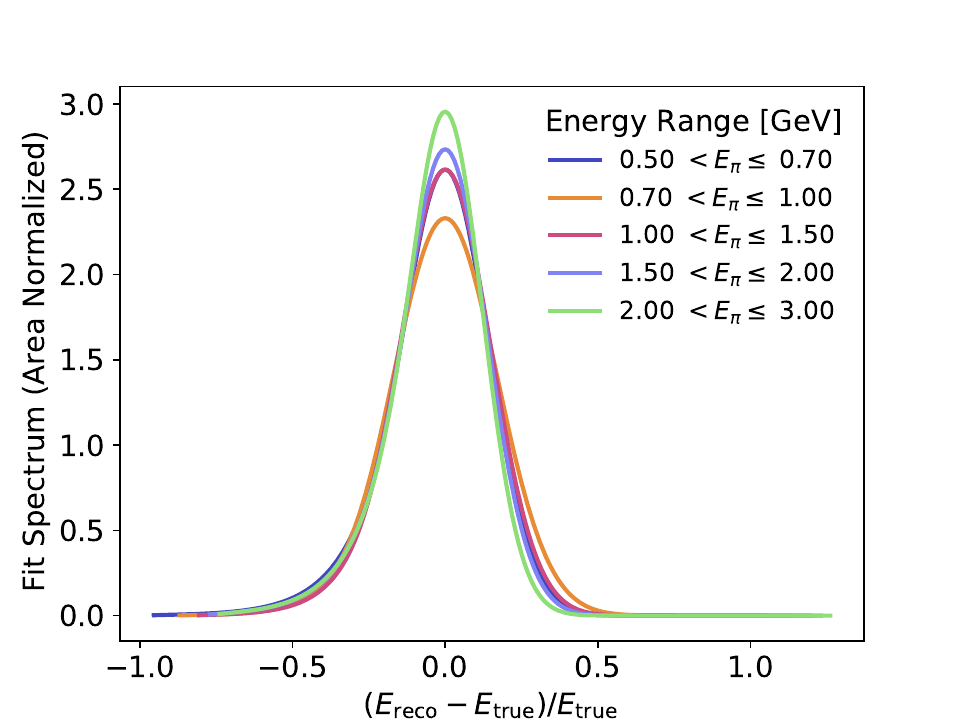}
    \caption{Fractional difference between the reconstructed pion energy and the true pion energy from  a GEANT4 simulation. The deposited energy from the pion in a DUNE ND-sized liquid argon volume is summed and smeared by 5\% to account for the caloriemtric energy resolution. The distribution in each pion energy bin is fit to a Crystal-ball function, and de-biased so that the peak is at 0. The de-biased fit is plotted for each pion energy range. }
\label{fig:pi_E_res}
\end{figure}

Simulated kinematic variables are smeared in a realistic way to emulate reconstruction effects and account for the projected detector performance. In DUNE ND, muons will either range out inside the LArTPC or be caught by a magnetized muon spectrometer, so muon momenta are smeared with a 5\% resolution. To estimate the pion reconstruction performance, we simulate the propagation of charged pions in GEANT4~\cite{GEANT4} in a volume of liquid argon using the LArSoft framework~\cite{LArSoft}. The pions were generated in the defined fiducial volume, and the deposited energy in the full DUNE ND active volume was accumulated and smeared by 5\% to account for the calorimetric energy resolution demonstrated by LArTPCs~\cite{ICARUSEnergyScale}. The distribution of the de-biased fractional difference between the ``reconstructed'' and true pion energy is shown in Fig.~\ref{fig:pi_E_res} for a few ranges of pion energy. This distribution is largely independent of the true pion energy, so we apply the distribution of the lowest pion energy (as fit to a Crystal-ball function~\cite{CrystalBall}) across the full range. This fit obtains an 18\% resolution in the reconstructed pion energy, with a non-Gaussian tail extending to lower values, consistent with results from the MicroBooNE experiment~\cite{MicroBooNEpionres}. This resolution neglects the impact of dead regions inside the modularized liquid argon detector that comprise about 10\% of the total volume and will have to be understood for this measurement to be performed in practice~\cite{DUNETDR}. 

Although the $\CCCoh$ process makes up only a small fraction of the total neutrino cross section at the few-GeV neutrino energy range, its distinct final state signature enables a relatively simple event selection which yields a high signal purity. We perform a mock $\CCCoh$ event selection benchmarking the procedures outlined in previous cross section measurements~\cite{CohPixsecCHARM-II, CohPixsecMINERvA1, CohPixsecMINERvA2, CohPixsecMINERvA3, CohPixsecT2K, ArgoNeuT:2014uwh}. First, we require the reconstructed event to have exactly two minimally ionizing tracks; events with any $\pi^0$, $\gamma$, or a proton with a kinetic energy above \SI{20}{MeV} are rejected. We then require the event to have a reconstructed $|t|$ smaller than 0.03 GeV$^2$. We also place upper threshold cuts on the muon energy (\SI{5.5}{GeV}) and pion energy (\SI{3}{GeV}) to remove events that are too energetic to be reconstructed effectively at DUNE ND. We do not impose a lower threshold on the reconstructed muon and pion energy because the rate of such events is small and they have a minimal impact on the power on the constraint. Contamination from $\CCCoh$ arising interactions from the wrong-sign component of the flux ($\bar\nu$ for FHC) are not included as backgrounds. These represent a small fraction of the event rate, and are expected to be suppressed by the ability of the spectrometer to identify the sign of the muon. After the event selection, the remaining background is subdominant and comprises the muon-neutrino resonant single pion production, shallow/deep-inelastic scattering, and neutral current scattering. This mock event selection leaves out other possible background rejection tools, such as low energy ``blip" identification, that could further separate the $\CCCoh$ process from incoherent backgrounds~\cite{ArgoNeuTLowE, Blips}.

We also exclude other potential background sources that are expected to have a small (sub-percent) impact. For example, $K_S^0 \to \pi^+\pi^-$ vertices would be tagged by the hadronic activity at the neutrino interaction (since the $K_S^0$ decay length, \SI{2.7}{\centi\meter}, is short). The process where this is not possible, neutral current coherent kaon production, has a cross section more than two orders of magnitude smaller than that of $\CCCoh$ in the energy region of interest~\cite{Alvarez-Ruso:2012kmi}. While accounting for such rare backgrounds would be an important component of applying this technique on data in a real constraint, we do not expect them to affect the conclusions of this feasibility study.

In addition to background rejection cuts, restrictions on the pion kinematics are made to remain in a region where the Adler relation is valid and the $\CCCoh$ cross section is constrained by $\pi$-Ar elastic scatters. Events with a reconstructed pion kinetic energy less than \SI{300}{MeV} are removed to minimize the sign correction in the $\pi^+$-Ar cross section. We assume that any measurement of the $\pi$-Ar cross section will be limited to scattering angles above $5^\circ$, so we also cut out $\CCCoh$ events that are below the equivalent momentum transfer to the nucleus ($t$). We express this requirement in terms of the effective pion scattering angle $\theta_\pi^\text{eff}$, where
\begin{equation}
\cos\theta_\pi^\text{eff} = 1 - 2 t / p_\pi^2\,.
\label{eq:thpieff}
\end{equation}
We exclude events with a reconstructed effective scattering angle below $6^\circ$. Finally, events with reconstructed $\zeta$ $(\equiv E_\pi / \sqrt{Q^2})$ greater than 3.25 and $Q^2 >\,$\SI{0.2}{GeV\squared} are cut to reduce the impact of transverse currents that spoil the Adler relation. In the case of $T_\pi$, $\theta_\pi^\text{eff}$, and $\zeta$, the cuts on the reconstructed variables (\SI{300}{MeV}, $6^\circ$, and 3.25, respectively) are tighter than the areas of true phase space where the constraint is not satisfied (\SI{250}{MeV}, $5^\circ$, and 3, respectively). This reduces the fraction of events that smear from unconstrained into constrained regions of phase space at the reconstruction level. 

The above restrictions on $\theta_\pi^\text{eff}$ and $T_\pi$ are necessary for the DUNE-ND measurement of the $\pi$-Ar cross section (see section~\ref{sec:piArEventSelection} for more details). In principle, they are not needed for an external measurement of that cross section. Including these regions of phase space in a constraint may be challenging: $\theta_\pi < 5^\circ$ because measuring the elastic cross section at such low angles is a challenge in many experimental setups~\cite{C12PiEl1, C12PiEl2}, and $T_\pi <~$\SI{250}{MeV} because that region has a significant component from transverse currents (small $\zeta$). We therefore exclude these regions from both the ``DUNE-ND-only'' and ``external'' scenarios of the fit. However, exploring these regions may prove to add further constraining power beyond the result presented here. 

\begin{figure}[]
    \centering
    \includegraphics[width=0.49\textwidth]{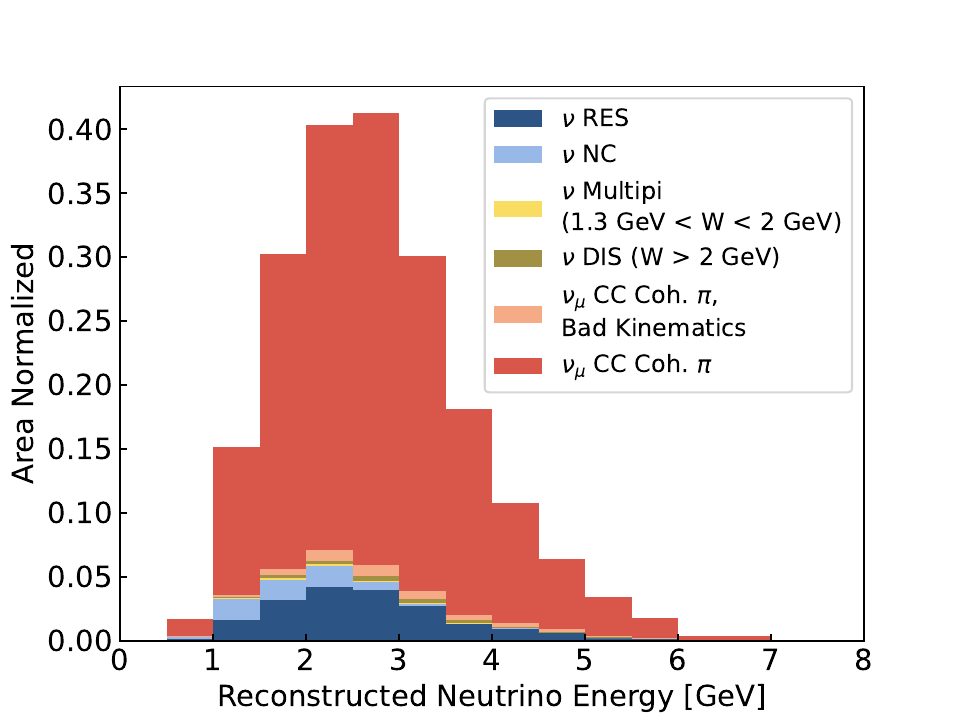}
    \caption{Distribution of reconstructed neutrino energy for simulated neutrino events in our DUNE ND simulation after topological and kinematic selection.}
\label{fig:cohpi_evt_sel}
\end{figure}

The reconstructed neutrino energy distribution of selected events is shown in Fig.~\ref{fig:cohpi_evt_sel}. This spectrum shows the distribution of selected $\CCCoh$ events against backgrounds from other neutrino interaction channels, as well as $\CCCoh$ with ``bad" kinematics, defined as events where: $T_\pi <$~\SI{250}{MeV}, $\theta_\pi^\text{eff} < 5^\circ$, $\zeta < 3$, or $Q^2 >$~\SI{0.2}{GeV\squared}. For these events, kinematics are such that either the Adler relation is not reliable, or the equivalent $\pi$-Ar elastic cross section cannot be reliably obtained from data.

\subsubsection{$\pi$-Ar Elastic Scattering Event Selection}
\label{sec:piArEventSelection}

\begin{figure}[]
    \centering
    \includegraphics[width=0.49\textwidth]{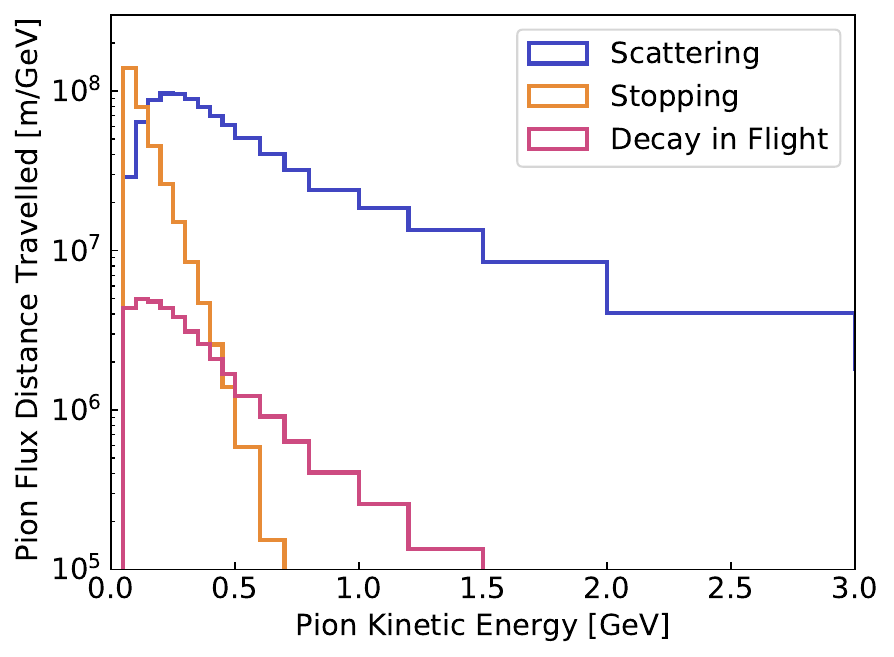}
    \caption{Distance traveled through the fiducial volume by pions produced in neutrino interactions in out DUNE ND simulation.}
\label{fig:pi_flux}
\end{figure}

Alongside the $\CCCoh$ event selection, a second mock event selection identifies $\pi$-Ar elastic scatters, to be used for the scenario where the cross section is simultaneously measured in DUNE ND. The neutrino-induced charged pion flux is found by propagating the pions from the GENIE simulation of neutrino interactions through a DUNE ND-sized volume of liquid argon in GEANT4~\cite{GEANT4} using the LArSoft framework~\cite{LArSoft}. The resulting flux, broken down by whether the pion stops, inelastically scatters, or decays inside the detector volume is shown in Fig.~\ref{fig:pi_flux}. 

\begin{figure}[]
    \centering
    \includegraphics[width=0.49\textwidth]{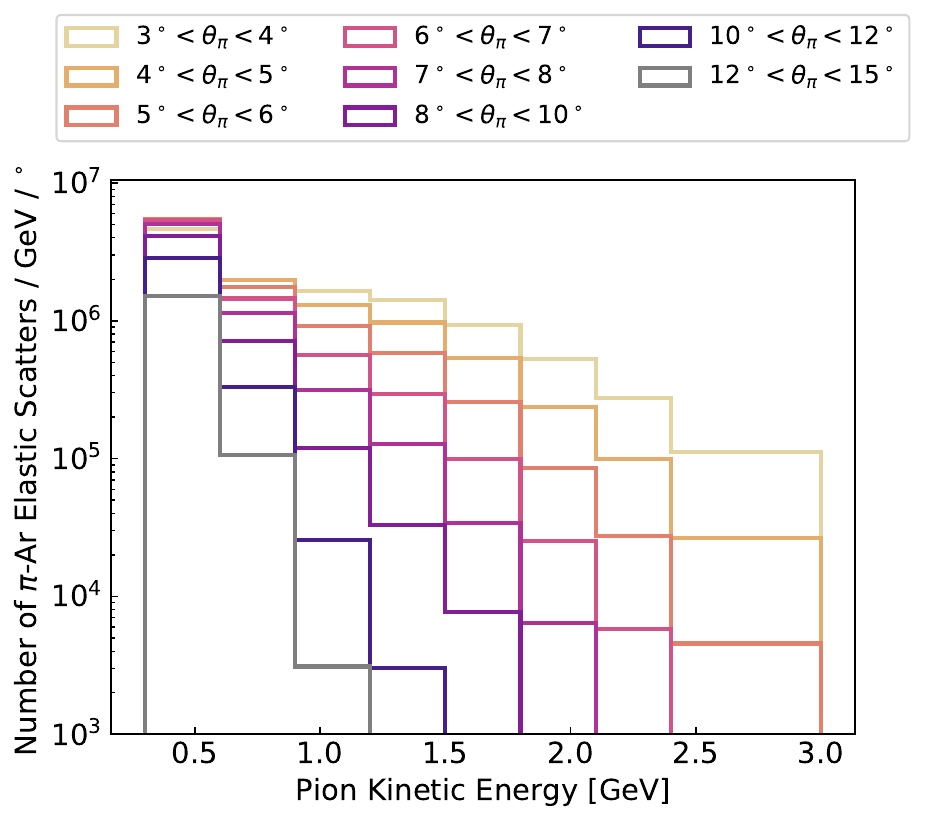}
    \caption{Pion elastic scatters from the neutrino induced pion flux at the DUNE ND, limited to pions that inelastically scatter in the detector.}
\label{fig:pi_elastic}
\end{figure}

The number of $\pi$-Ar elastic scatters is computed with the equivalent $\pi$-Ar elastic cross section model as used by the GENIE Berger-Sehgal/Rein-Sehgal simulation. Although this is a less sophisticated simulation of $\pi$-Ar scatters, using the same model for $\CCCoh$ scattering and $\pi$-Ar elastic scattering ensures consistency in the fit. The pion flux used includes only those pions that inelastically scatter, for which there should be no significant background from other tracks. Protons can be separated from pions calorimetrically, and muons do not inelastically scatter before stopping. The energy-angle distribution of $\pi$-Ar elastic scatters is shown in Fig.~\ref{fig:pi_elastic}.

Backgrounds to elastic scatter vertices should be insignificant. The LArIAT experiment found that the background from multiple-Coulomb-scattering is small at the angle threshold used in this analysis ($\theta_\pi > 5^\circ$)~\cite{LArIATxsec}. We found the background from soft inelastic scatters from the propagation of neutrino-induced pions through a DUNE ND-sized volume of liquid argon as simulated by GEANT4. We considered a fake elastic scatter as any inelastic process that produces a single charged pion in the final state, and no protons, photons, or $\pi^0s$ above the same energy thresholds used in the neutrino selection (\SI{20}{MeV} KE). Background scattering events are dominated by cases where only high energy neutrons are produced in the final state. The distribution of the apparent scattering angle (given as the angle between the ingoing and outgoing pion momenta) is much broader than the equivalent elastic scattering distribution, peaking at about $\theta_\pi^\mathrm{apparent} \approx 35^\circ$. The number of pions that produce a fake elastic scatter in the angle region of interest ($\theta_\pi < 15^\circ$) is very small: about 1 in 10,000. This translates to a few thousand fake elastic scatters for 5 years of running at DUNE, compared to the millions of elastic scatters. Although taking this background into account will likely be important for performing the measurement in practice, it should not significantly impact the result. We therefore neglect the background in the present analysis.

\begin{figure}[]
    \centering
    \includegraphics[width=0.49\textwidth]{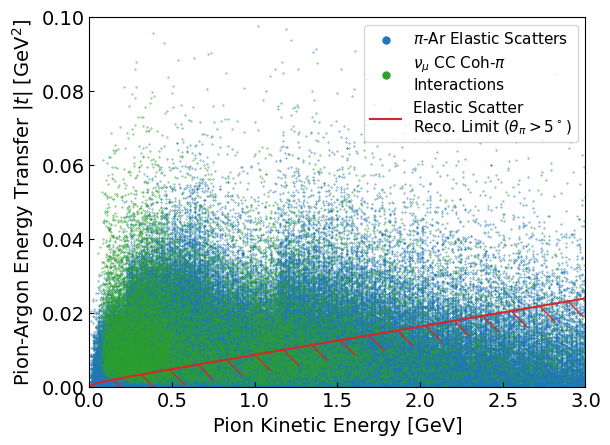}
    \caption{Distribution of pion kinetic energy and pion to argon energy transfer ($|t|$) for pions from $\CCCoh$ events (green) and $\pi$-Ar elastic scattering (blue) events in DUNE ND. Red hatched lines represent the cut on $\pi$-Ar scattering angle from limitations on reconstruction.}
\label{fig:nu_pi_overlap}
\end{figure}

The pion kinetic energy is smeared by the same kernel as for the neutrino event selection, and no experimental resolution is applied for the scattering angle. The intrinsic angular resolution for  particle tracks in a LArTPC is very precise ($\sim$\SI{3}{\milli\radian}~\cite{MicroBooNE:2017tkp}), but in a real measurement, care would have to be taken to minimize the impact of multiple-Colomb-scattering. The Highland formula~\cite{MCS1,MCS2} predicts a scattering width of \SI{6}{\milli\radian} in LAr across one pixel pitch (projected to be \SI{4}{\milli\meter} in DUNE-ND~\cite{DUNENDCDR}) at the lowest pion kinetic energy considered in our analysis (\SI{250}{MeV}). This width falls to \SI{2}{\milli\radian} at a pion kinetic energy of \SI{1}{GeV}. Finally, $\pi$-Ar elastic scatters are required to be at an angle greater than $5^\circ$, in accordance with the threshold achieved by the LArIAT experiment~\cite{LArIATxsec}. For the flux constraint to be feasible, a sufficient overlap of the kinematic phase space covered by the selected $\CCCoh$ sample and the $\pi$-Ar elastic scattering sample is essential. This coverage is illustrated in Fig.~\ref{fig:nu_pi_overlap}.

Although we take the $5^\circ$ reconstruction threshold as fixed for this analysis, in principle future measurements of the $\pi-$Ar cross section (not performed with LArTPCs) could be sensitive to lower angles. Expanding to lower pion scattering angles would correspond to allowing in more $\uCCCoh$ events at low energy transfer ($|t|$). This would increase the number of events available to the analysis, in a kinematic region where systematic effects degrading the interaction coherence should be smallest.
\subsection{Flux Constraint}\label{sec:flux_constraint}
\begin{figure*}[]
    \centering
    \subfloat[\label{sfig:nutemp_1}]{
    \includegraphics[width=0.24\linewidth]{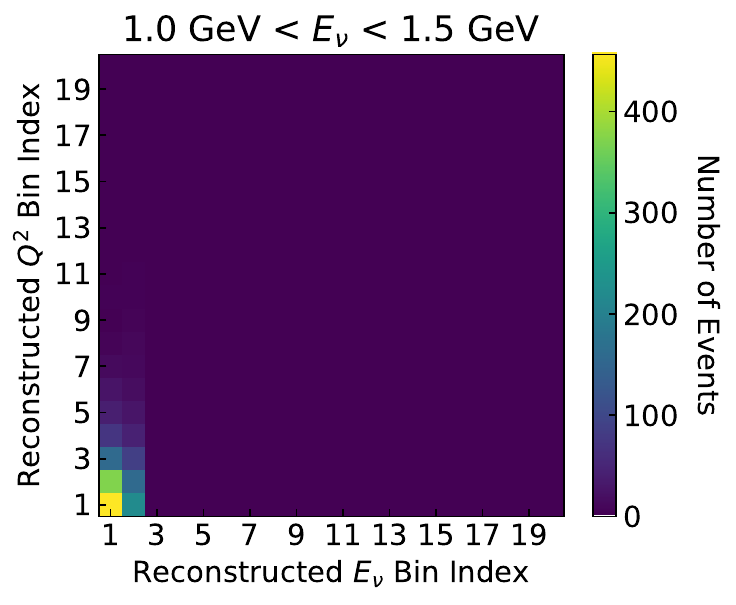}}
    \subfloat[\label{sfig:nutemp_2}]{
    \includegraphics[width=0.24\linewidth]{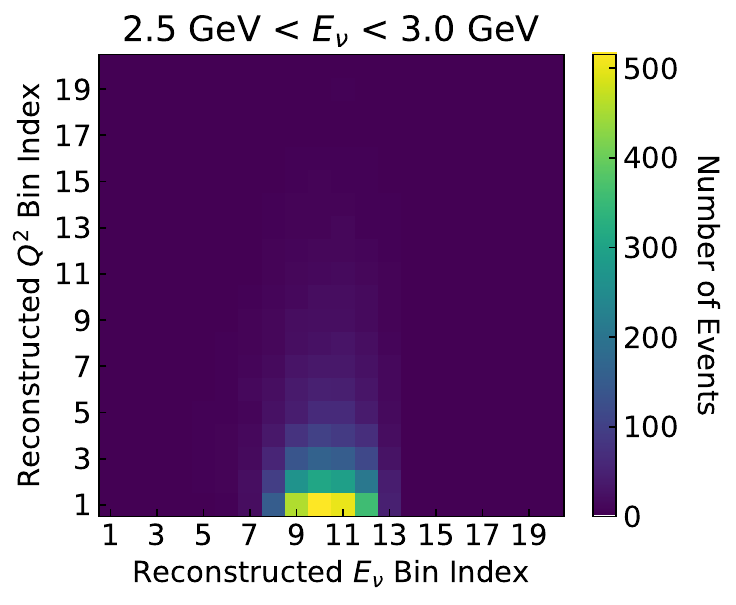}}
    \subfloat[\label{sfig:nutemp_3}]{
    \includegraphics[width=0.24\linewidth]{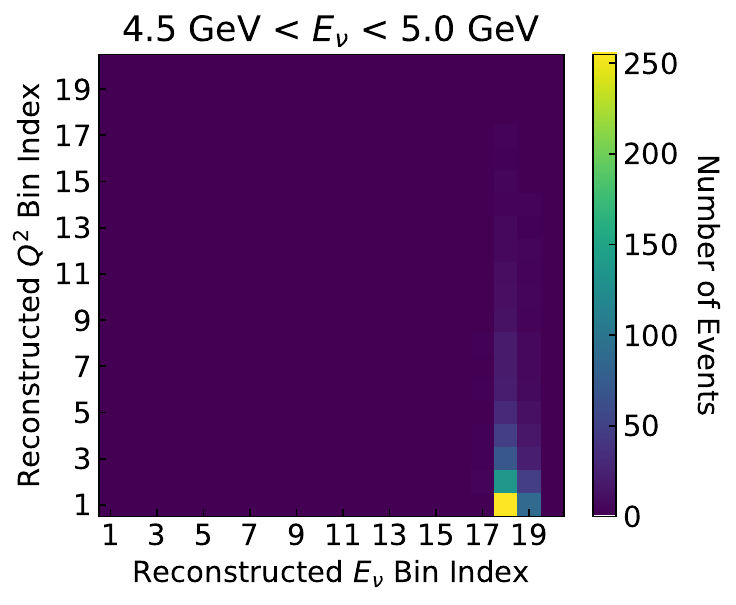}}
    \subfloat[\label{sfig:bgtemp}]{
    \includegraphics[width=0.24\linewidth]{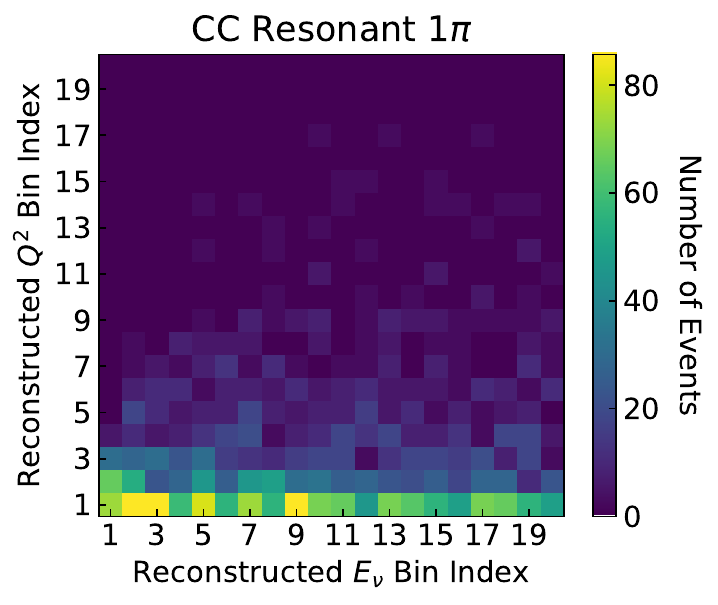}}
    \caption{Example fit templates of $\CCCoh$ interaction events for different true neutrino energy ranges of (a) 1.0-1.5$\,$GeV (b) 2.5-3.0$\,$GeV (c) 4.0-5.0$\,$GeV and (d) for the leading background interaction mode, charged-current resonant single pion production.}
    \label{fig:templates}
\end{figure*}

The 2D distribution of reconstructed neutrino energy and reconstructed $Q^2$ of simulated $\CCCoh$ events are divided into neutrino energy bins. The distribution for each bin serves as the template for the corresponding true neutrino energy range in the fit procedure outlined in this section. In addition, templates for the 4 leading background modes are included in the fit to account for imperfect event selection purity. Figure~\ref{fig:templates} shows examples of $\CCCoh$ and background templates. The binning used to construct these templates is listed in Table~\ref{table:bins}. Linear bins were chosen for reconstructed $Q^2$, while reconstructed neutrino energy bins were chosen so that each bin would have the same number of events.

\begin{table}[]
    \centering
    \begin{tabular}{|p{2.4cm}|p{5.8cm}|} 
     \hline
     Template \hspace{15pt} $E_{\nu}$ Bins [GeV] & 0, 0.5, 1.0, 1.5, 2.0, 2.5, 3.0, 3.5, 4.0, 4.5, 5.0, 5.5, 6.0, 7.0, 8.0, 12.0, 16.0, 20.0, 40, 100  \\
     \hline
     \hline
     Reconstructed $Q^2$ Bins [GeV$^2$] & 0, 0.01, 0.02, 0.03, 0.04, 0.05, 0.06, 0.07, 0.08, 0.09, 0.1 ,
       0.11, 0.12, 0.13, 0.14, 0.15, 0.16, 0.17, 0.18, 0.19, 0.2 \\
     \hline 
     Reconstructed $E_{\nu}$ Bins [GeV]& 0.00, 1.37, 1.63, 1.82, 1.98, 2.12, 2.26, 2.39, 2.52, 2.64, 2.76, 2.89, 3.03, 3.19, 3.36, 3.57, 3.84, 4.22, 4.86, 7.34, 100.00 \\
     \hline
    \end{tabular}
    \caption{Summary of choice of bins for neutrino event templates. The true neutrino energy bins for template assignment are taken from the binning of the flux covariance matrix of Fig.~\ref{fig:flux_cov_cor}. Reconstructed $Q^2$ bins are linear, while reconstructed neutrino energy bins make the number of events in each reconstructed neutrino energy bin even.}
    \label{table:bins}
\end{table}

The template fit is executed by minimizing the log-likelihood function given by the combined chi-squared
\begin{equation}
     \chi^2 = \left(\chi_{\nu}^\text{stat}\right)^2 + \left(\chi_{\pi}^\text{stat}\right)^2 + \left(\chi_{\pi-\text{Ar}}^\text{prior}\right)^2 +\left(\chi_{\mathcal{F_A}}^\text{prior}\right)^2 \,.
\end{equation}
The four terms represent the four components of the fit: the $\CCCoh$ data $\left(\chi_{\nu}^\text{stat}\right)^2$, the $\pi$-Ar elastic scatter data for the DUNE ND only scenario  $\left(\chi_{\pi}^\text{stat}\right)^2$, the prior on the $\pi$-Ar elastic cross section $\left(\chi_{\pi-\text{Ar}}^\text{prior}\right)^2$, and the prior on the form factor $\left(\chi_{F_{A}}^\text{prior}\right)^2$. The $\left(\chi_{\nu}^\text{stat}\right)^2$ term is the binned log-likelihood for the neutrino events,
\begin{equation}
\begin{split}
    \chi_{\nu}^2 = {} & 2\sum_{i=1}^N \mu_i (\vec{x}_\nu, \vec{x}_{\pi-Ar}, \vec{x}'_{\pi-Ar}, \vec{a}) \\ 
    & - n_i +n_i \mathrm{log} \frac{n_i}{\mu_i (\vec{x}_\nu, \vec{x}_{\pi-Ar}, \vec{x}'_{\pi-Ar}, \vec{a})},
\end{split}
\end{equation}
where $n_i$ is the number of data events in bin $i$, $\vec{x}_\nu$ is the vector of neutrino template normalization parameters, and $\vec{x}_{\pi-Ar}$, $\vec{x}'_{\pi-Ar}$, and $\vec{a}$ are parameters that determine the $\CCCoh$ cross section entering the event rate calculation ($\mu_i$). The rate $\mu_i$ is given as the weighted number of simulation events in bin $i$. The weight on each background event is equal to the template scale factor of its background category (resonant pion production, neutral current scattering, multiple pion production, or deep inelastic scattering). The weight $w_{\CCCoh}$ on each $\CCCoh$ event is 

\begin{widetext}
\begin{equation}
    w_{\CCCoh} (E_\nu, Q^2, T_\pi, \theta_\pi^\text{eff}) = \vec{x}_\nu(E_\nu) \times \frac{\xi_{\rm Adler}^{\rm z-exp}(\vec{a}, Q^2)}{\xi_{\rm Adler}(Q^2)} \times \begin{cases}
    \vec{x}_{\pi-Ar}(T_\pi, \theta_\pi^\text{eff}), & \text{if } \zeta > 3 \text{ and } T_\pi > \text{250 MeV} \\
    \vec{x}'_{\pi-Ar}(T_\pi, \theta_\pi^\text{eff}),              & \text{otherwise}
\end{cases}\,.
\end{equation}
\end{widetext}
In this equation, $\xi_{\rm Adler}^{\rm z-exp}(\vec{a}, Q^2)$ is the Adler screening factor (Eq.~\ref{eq:cs_adler_3}) with the $z$-expansion axial form factor determined by the fit parameters $\vec{a}$ (Eq.~\ref{eq:zexpFF}) and $\xi_{\rm Adler}(Q^2)$ is the screening factor with the initial form factor in the simulation. The $z$-expansion form factor is modeled with 4 coefficients, and the $t_0$ value is chosen as -0.07, to minimize the mean value of $|z|$ for the $Q^2$ distribution of the simulated $\CCCoh$ neutrino events. The neutrino flux template scale factor $\vec{x}_\nu$ is a function of the true neutrino energy. The $\pi$-Ar scale factors for different phase space are noted with $\vec{x}_{\pi-Ar}$ and $\vec{x}'_{\pi-Ar}$. $\CCCoh$ events in a region of parameter space that can be constrained by the Adler relation, as defined by $\zeta > 3 \text{ and } T_\pi > \text{250 MeV}$, are fit to the $\vec{x}_{\pi-Ar}$ scale factors. As will be detailed below, the $\vec{x}_{\pi-Ar}$ scale factors are constrained by the $\pi$-Ar part of the fit. The identification of these scale factors between the pion and neutrino data enables the $\pi$-Ar measurement to constrain the $\CCCoh$ event rate in the fit. $\CCCoh$ events with bad kinematics that are included in the sample are fit to the second set of $\pi$-Ar scale factors ($\vec{x}'_{\pi-Ar}$) that are not constrained by the $\pi$-Ar elastic scattering cross section. In the DUNE ND only scenario of the fit, the $\pi$-Ar scale factors are binned by the true pion kinetic energy and the effective scattering angle (Eq. \ref{eq:thpieff}). In the external $\pi$-Ar scattering data scenario, the $\pi$-Ar scale factor is given by a single number used for all events.

\begin{table}[t]
    \centering
    \begin{tabular}{ |l|p{4cm}| } 
     \hline
     Template $T_{\pi}$ Bins [GeV]& 0, 0.3, 0.6, 0.9, 1.2, 1.5, 1.8, 2.1, 2.4, 3, 10 \\ 
     \hline
    Template $\theta_{\pi}$ Bins& $0^\circ$, $2^\circ$, $3^\circ$, $4^\circ$, $5^\circ$, $6^\circ$, $7^\circ$, $8^\circ$, $10^\circ$, $12^\circ$, $15^\circ$, $180^\circ$ \\ 
     \hline 
     \hline
     Reconstructed $T_{\pi}$ Bins [GeV]&  0.3, 0.6, 0.9, 1.2, 1.5, 1.8, 2.1, 2.4, 3, 10 \\ 
     \hline
    Reconstructed $\theta_{\pi}$ Bins& $5^\circ$, $6^\circ$, $7^\circ$, $8^\circ$, $10^\circ$, $12^\circ$, $15^\circ$, $180^\circ$ \\
     \hline
    \end{tabular}
    \caption{Summary of choice of bins for pion scatter templates. The template bins and the reconstructed bins are the same, except for a cut on the kinetic energy of \SI{0.3}{GeV} and on the scattering angle of $5^\circ$.}
    \label{table:pibins}
\end{table}

The $\left(\chi_{\pi}^\text{stat}\right)^2$ term is included in the DUNE ND only scenario. In this case, $\pi$-Ar elastic scattering events in the fit are binned by the reconstructed pion kinetic energy and the $\pi$-Ar scattering angle. They are split into templates as a function of the true pion kinetic energy and $\pi$-Ar scattering angle. The $\pi$-Ar elastic scattering cross section is fit for as scale factors on each of these templates. Both the template and reconstructed binning is detailed in Table~\ref{table:pibins}. The $\left(\chi_{\pi-\text{Ar}}^\text{stat}\right)^2$ term is computed as the binned log-likelihood for pion scattering events,
\begin{equation}
    \chi_{\pi}^2 = 2\sum_{i=1}^N \left[ \mu_i (\vec{x}_{\pi-Ar}) - n_i +n_i \mathrm{log} \frac{n_i}{\mu_i (\vec{x}_{\pi-Ar})} \right] ,
\end{equation}
where $\mu_i (\vec{x}_{\pi-Ar})$ is the weighted number of simulation events in bin $i$, $\vec{x}_{\pi-Ar}$ is the vector of $\pi$-Ar elastic scattering template scale factors, and $n_i$ is the number of data events in bin $i$. The weight on each $\pi$-Ar scattering event in each template $j$ is equal to the scale factor on that template: ($\vec{x}_{\pi-Ar})_j$. 

Term $\left(\chi_{\pi-Ar}^\text{prior}\right)^2$ is the prior on the two sets of $\pi$-Ar cross section weights. It is equal to
\begin{equation}
    \chi^2_{\pi-\text{Ar}} = \left|\frac{1 - \vec{x}_{\pi-Ar}}{\sigma^{\rm prior}_{\pi}}\right|^2 + \left|\frac{1 - \vec{x}'_{\pi-Ar}}{0.3}\right|^2\,,
\end{equation}
where $\sigma^{\rm prior}_{\pi}$ is the prior uncertainty on the $\pi$-Ar cross section. For the external $\pi$-Ar scattering data scenario, we vary the size of  $\sigma^{\rm prior}_{\pi}$ to study how the power of the external measurement impacts the flux constraint. For the DUNE ND only scenario, $\sigma^{\rm prior}_{\pi}$ is taken to be 30\%. In this case, the prior is loose, so in any bin where there is coverage from $\pi$-Ar scattering data, the statistical power of the data dominates. The prior acts to constrain scale factors that are not covered by $\pi$-Ar elastic scattering data (such as events with bad kinematics) so that the corresponding scale factor fit is not unrealistic. Only a small fraction of the neutrinos have such scale factors, so the precise value of the prior does not impact the result of the fit. 

Finally, we use the well predicted value of the form factor at $Q^2=0$ as a Gaussian prior
\begin{equation}
    \left(\chi_{\mathcal{F_A}}^\text{prior}\right)^2 = -0.5 \frac{(\mathcal{F_A}(\vec{a}, Q^2=0)-\mu_{\mathcal{F_A}(0)})^2}{\sigma_{\mathcal{F_A}(0)}^2}\,,
\end{equation}
where $\mu_{\mathcal{F_A}(0)}$ is known as unity up to the uncertainty $\sigma_{\mathcal{F_A}(0)}$ ($=$0.2$\%$), from calculation based on the uncertainty on the values of the pion decay constant and the Fermi constant~\cite{PDG}. This prior pins the strength of the cross section at $Q^2=0$ to the Adler relation.

In total in the DUNE ND only case, there are 247 parameters in the fit: 4 template scale factors on the neutrino background (one per channel), 19 $\CCCoh$ scale factors (one per true neutrino energy bin, see Table~\ref{table:bins}), 4 form factor parameters (one for each component in the z-expansion), and 220 pion scale factors. The pion scale factors are per-$T_\pi$ and per-$\theta_\pi$ bin (see Table~\ref{table:pibins}), with one copy each for $\uCCCoh$ events with and without good kinematics: $10\times 11\times 2 = 220$ parameters. In the case where we assume external $\pi$-Ar elastic scattering data, there are 29 total parameters. In this case there are only two total pion scale factors, one each for good and bad kinematics.

We use the \texttt{emcee}~\cite{EMCEE} package, a Python implementation of Goodman $\&$ Weare’s Affine Invariant Markov chain Monte Carlo (MCMC) Ensemble sampler~\cite{AffineMCMC} to fit the simulation to data by numerically maximizing the likelihood function. Running the MCMC chain enables us to study the posterior distribution and the covariance between the varied parameters. 

\section{Results}
\label{sec:result}
\subsection{Template Fit}
\begin{figure*}[]
    \centering
    \includegraphics[width=0.7\textwidth]{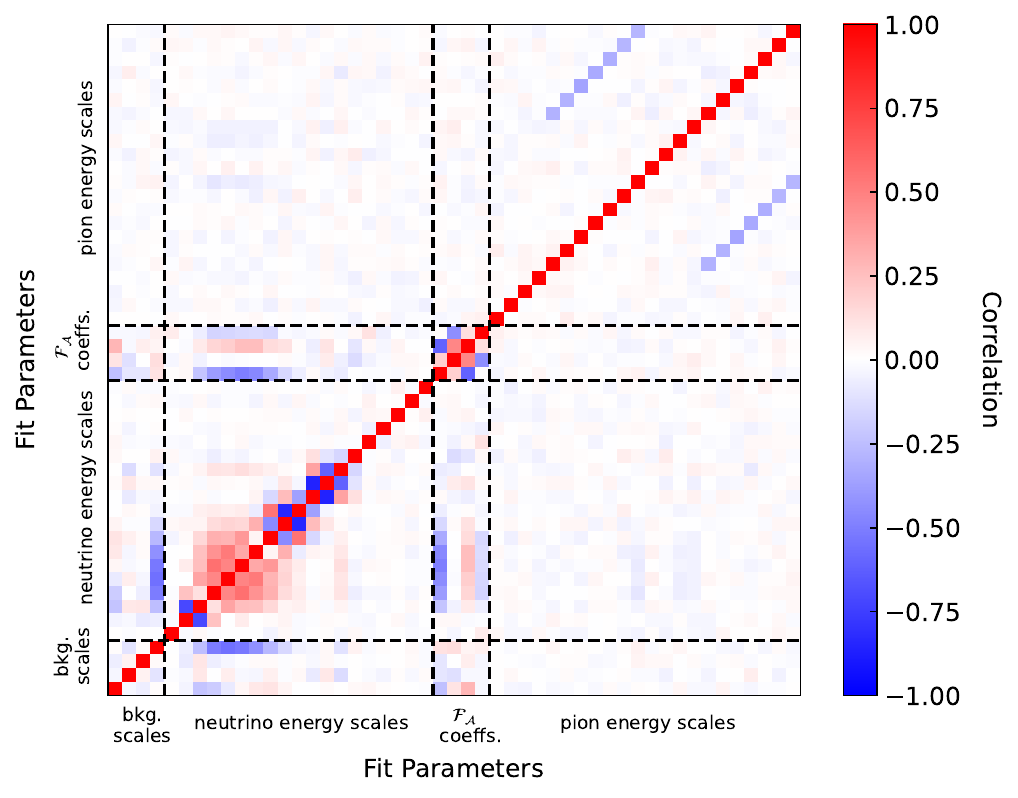}
    \caption{Correlation matrix between the fit parameters: 4 background template scale factors, 19 $\CCCoh$ template scale factors, 4 form factor parameters, and the first 22 pion cross section scale factors.}
\label{fig:fit_cov}
\end{figure*}

The covariance and correlation between fit parameters are obtained from the MCMC chain. Fig.~\ref{fig:fit_cov} shows the correlation between the fit parameters for the DUNE ND only scenario, up to the first 22 of the pion cross section scale factors corresponding to the first two pion angle bins. The remaining pion scale factor parameters follow patterns similar to the ones shown in the figure. In the following subsections, we detail the fit result for the form factor, the pion scale factors, and finally the neutrino flux scale factors.

\subsubsection{Form Factor}

\begin{figure}[]
    \includegraphics[width=\linewidth]{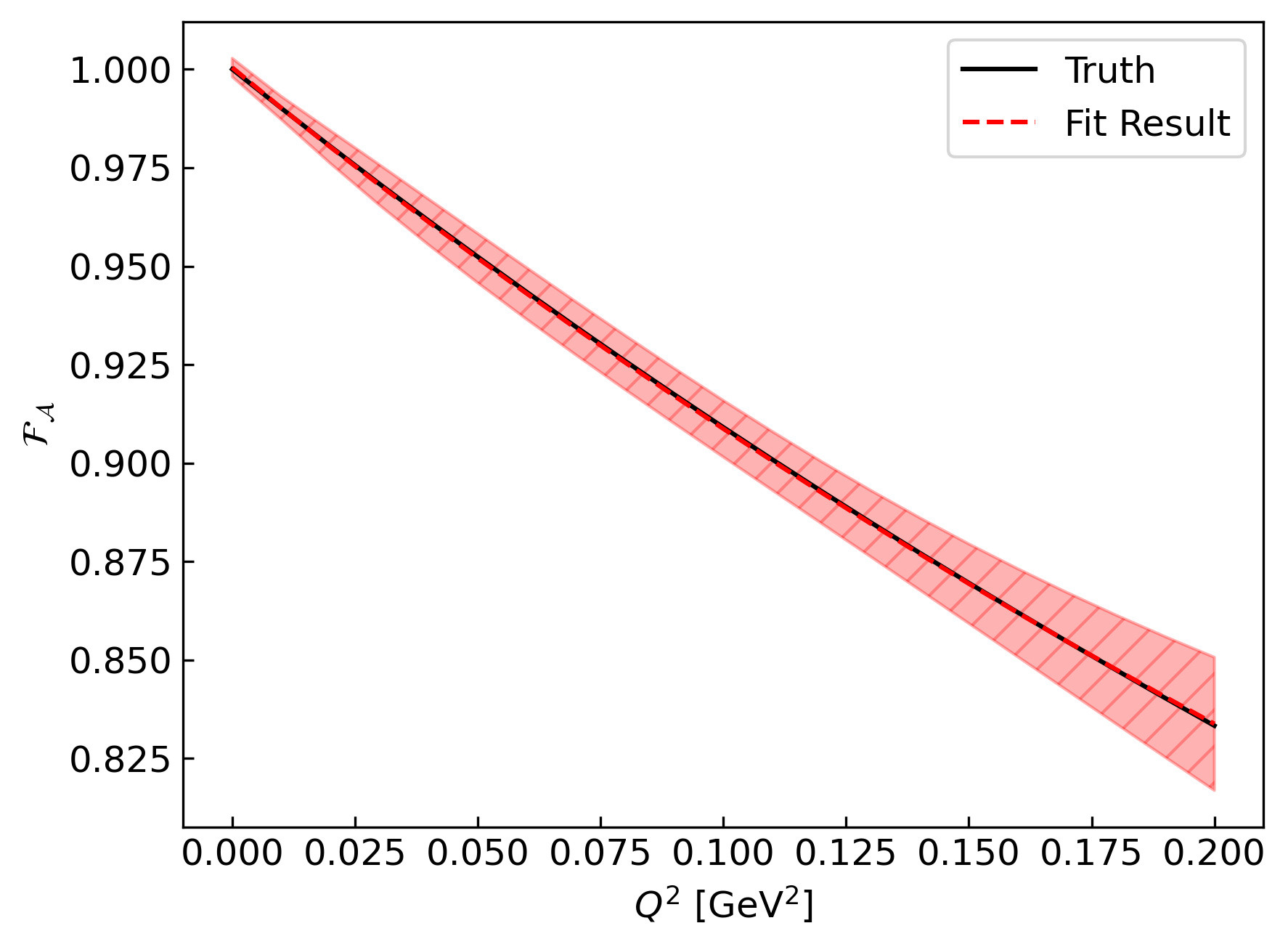}
    \caption{The axial-vector form factor fit result compared to the GENIE simulation truth. The hatched area spans the fit uncertainty.}
\label{fig:ff_fit}
\end{figure}

Figure~\ref{fig:ff_fit} shows the fitted form factor as a function of $Q^2$ compared to the GENIE simulation truth for the DUNE ND only scenario. The fit shows good agreement with the truth, with the uncertainties fully spanning the residual discrepancy for all $Q^2$ values. This demonstrates that the fit method does not introduce bias due to the correlation between the form factor coefficients and the neutrino energy scales, and that the $z$-expansion parameterization is capable of describing the $Q^2$ dependence in a model-independent way. The fit is strongly constrained at lower $Q^2$ values due to the prior imposed on the value at $Q^2=0$, and the uncertainty increases at higher $Q^2$ values.

\subsubsection{Pion-Argon Elastic Scattering Scale Factors}
\label{subsec:PiArConstraint}

\begin{figure*}[]
    \centering
    \subfloat[\label{sfig:pixsec_unc}]{
    \includegraphics[width=0.49\linewidth]{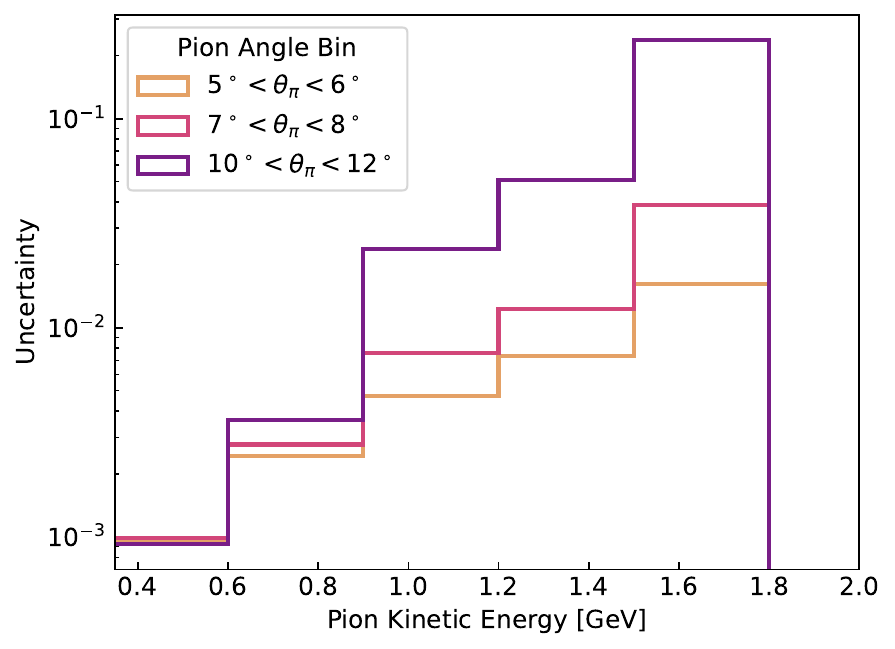}}
    \subfloat[\label{sfig:pixsec_corr}]{
    \includegraphics[width=0.45\linewidth]{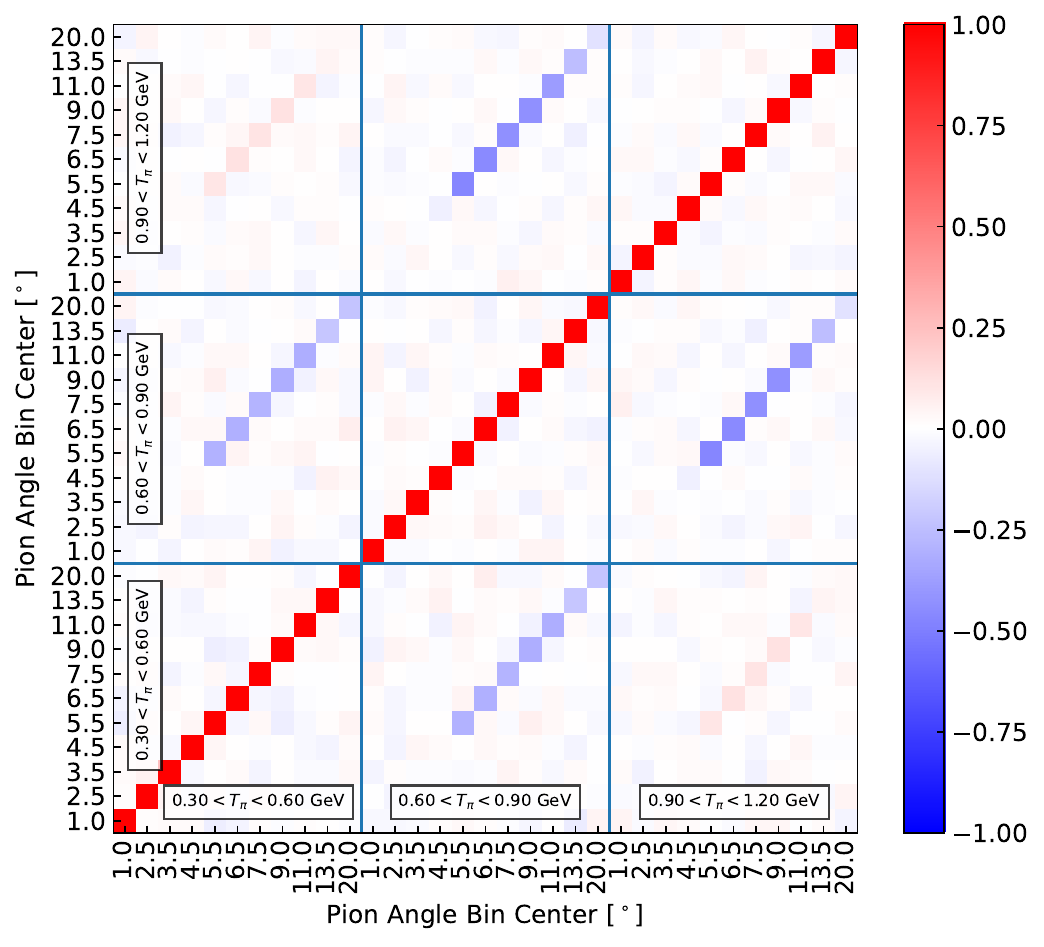}}
    \caption{(a) The uncertainty in the postfit value of the $\pi$-Ar elastic cross section weights shown as a function of the pion kinetic energy for a few scattering angle bins. (b) The post-fit correlation matrix of $\pi$-Ar elastic cross section weights, shown for all angle bins of the first 3 kinetic energy bins.}
\label{fig:pixsec_fit}
\end{figure*}

\begin{figure*}[]
    \centering
    \subfloat[\label{sfig:pixsec_nubin_unc}]{
    \includegraphics[width=0.48\linewidth]{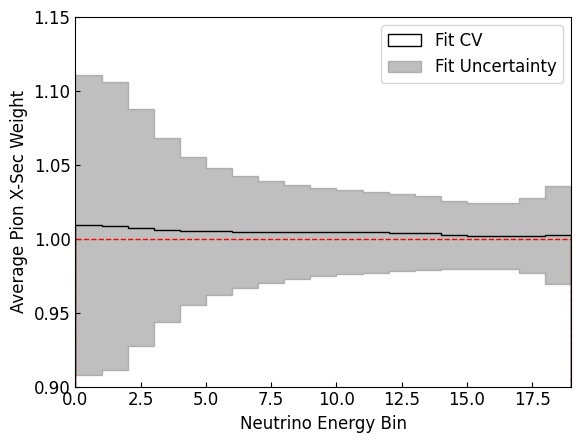}}
    \subfloat[\label{sfig:pixsec_nubin_corr}]{
    \includegraphics[width=0.48\linewidth]{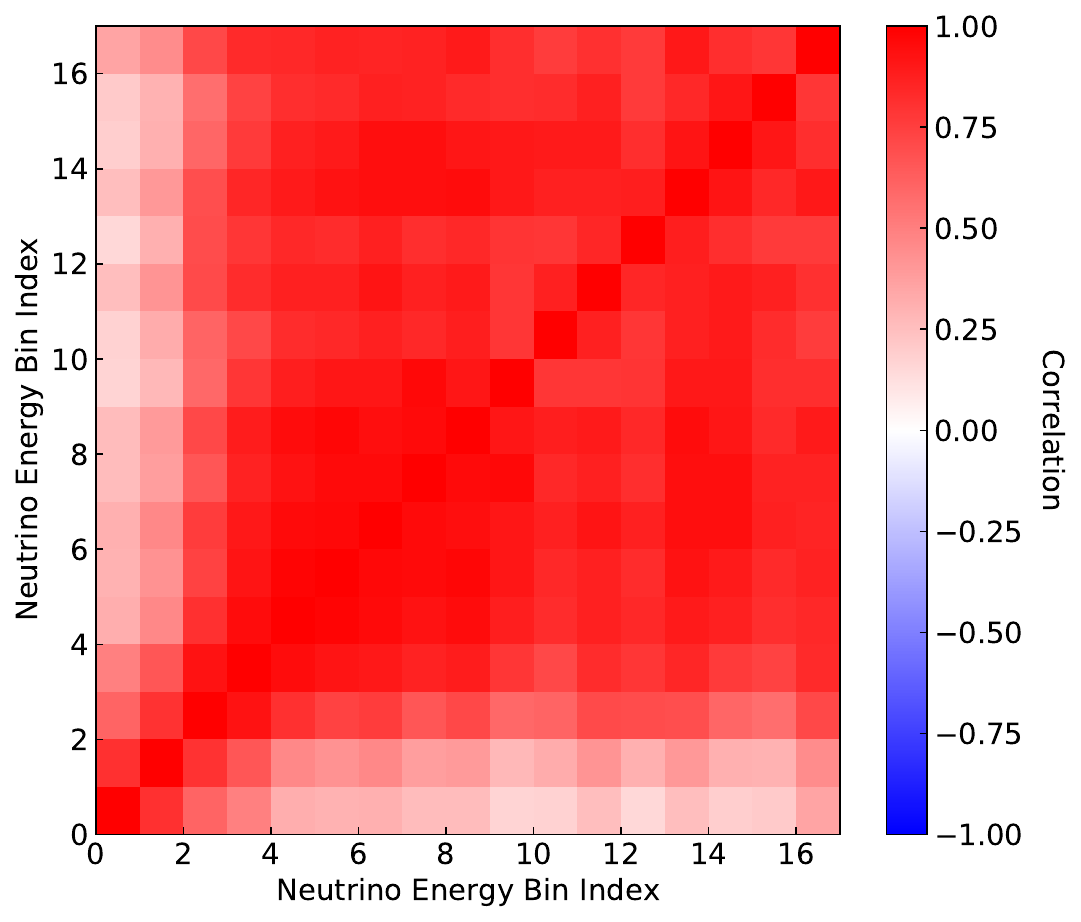}}
    \caption{(a) Post-fit uncertainties on the $\pi$-Ar elastic cross section weights, averaged over the $\CCCoh$ events in each neutrino energy bin. (b) The correlation matrix between different true neutrino energy bins. See Table \ref{table:bins} for the value of the binning.}
\label{fig:pixsec_nu_fit}
\end{figure*}

As shown in Fig.~\ref{fig:pixsec_fit}, the large number of $\pi$-Ar elastic scatters ($\mathcal{O}(10^7)$) available in a DUNE ND measurement enables the cross section on the process to be constrained to the sub-percent level. Some of that uncertainty is the result of the resolution in the pion energy measurement, which is visible in the off-diagonal correlations in the bottom plot of Fig.~\ref{fig:pixsec_fit}. Figure~\ref{fig:pixsec_nu_fit} demonstrates how the $\pi$-Ar part of the fit impacts the $\CCCoh$ part. The figure shows the uncertainty of the $\pi$-Ar elastic cross section weight averaged over all of the $\CCCoh$ events in each neutrino energy bin (see Table~\ref{table:bins} for the binning). The averaged cross section is constrained to the sub-percent level in the peak of the neutrino energy flux. The uncertainty is largely correlated between the different neutrino bins. This is because most of the neutrino bins have a largely overlapping pion kinematic phase space, especially after the kinematic restrictions in the event selection.

Since the statistical power is so strong, systematic uncertainties will also necessarily play a role in the application of this technique at DUNE. One source of possible systematic uncertainty is the correction of the sign-averaged pion cross section to the $\pi^+$ cross section relevant for $\CCCoh$ scattering. Another is any systematic uncertainty associated with detector performance, although these could also be reduced in the ratio of the $\CCCoh$ cross section to the $\pi$-Ar elastic scattering cross section. 

\begin{figure*}[]
    \centering
    \subfloat[\label{sfig:pre_cov}]{
    \includegraphics[width=0.49\linewidth]{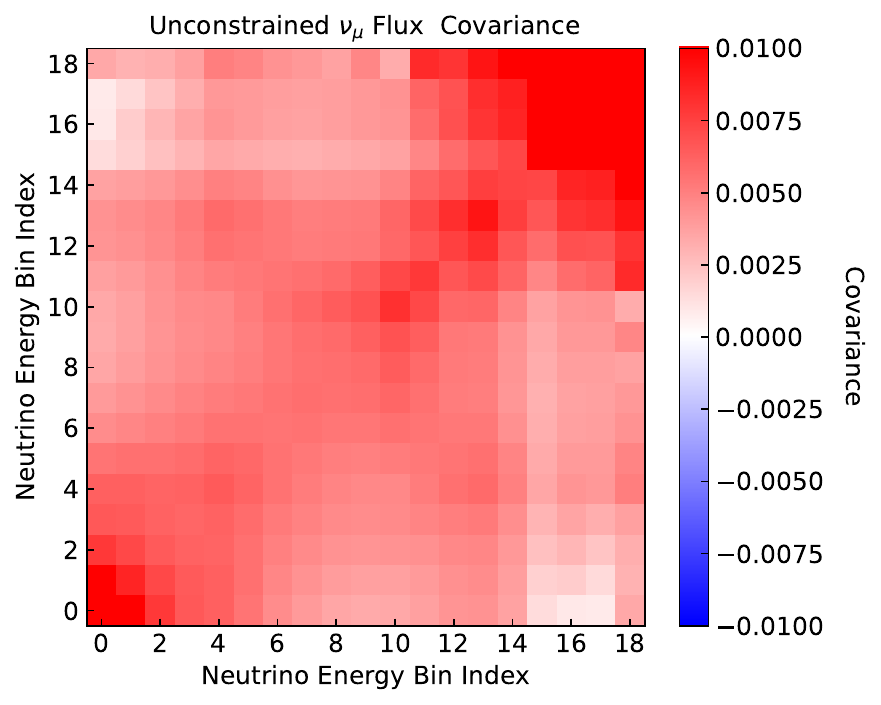}}
    \subfloat[\label{sfig:post_cov}]{
    \includegraphics[width=0.49\linewidth]{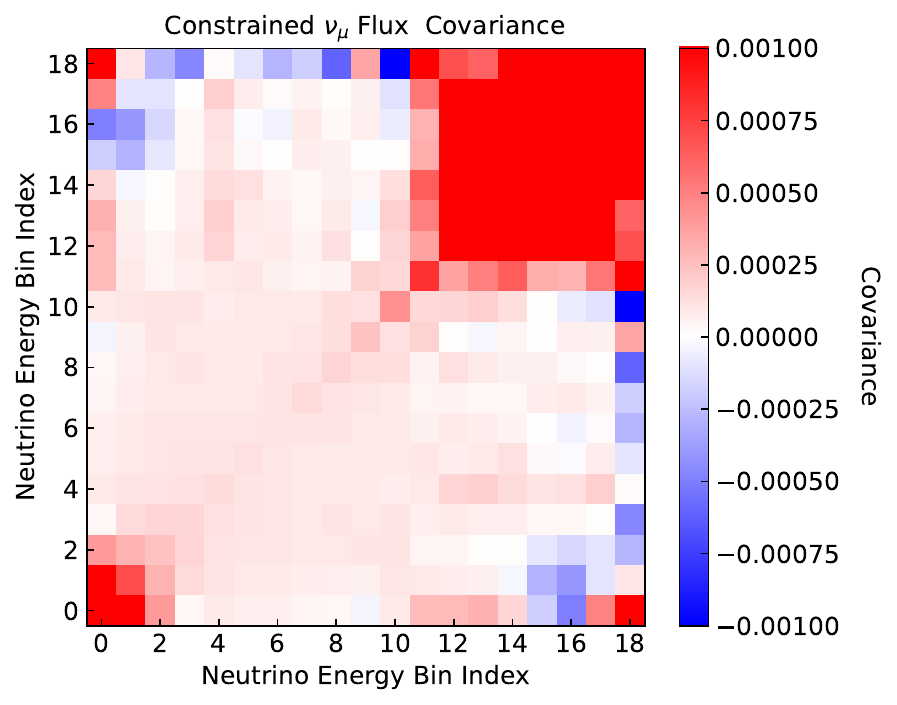}}
    \\ 
    \subfloat[\label{sfig:pre_corr}]{
    \includegraphics[width=0.48\linewidth]{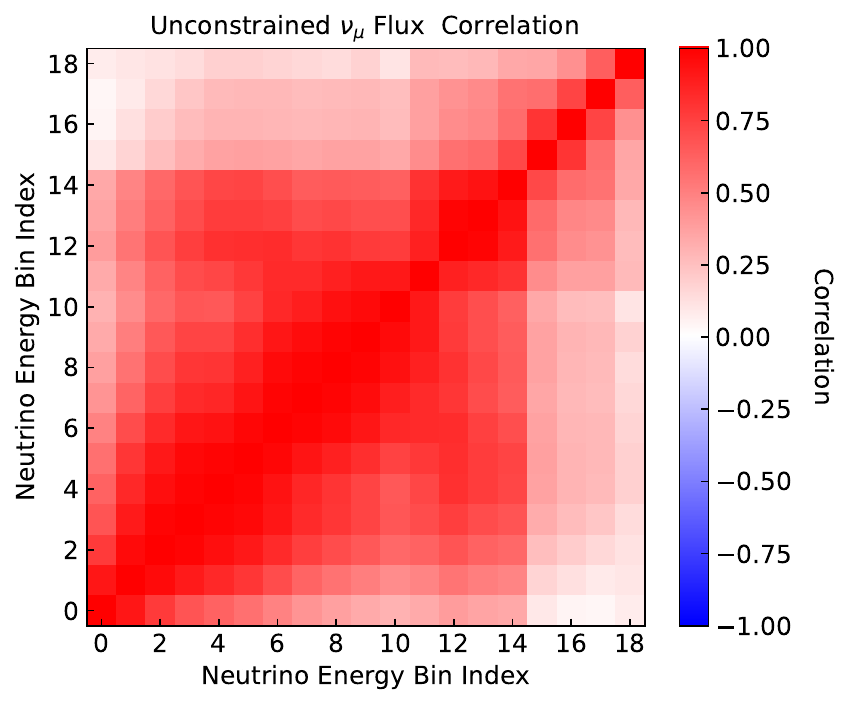}}
    \subfloat[\label{sfig:post_corr}]{
    \includegraphics[width=0.48\linewidth]{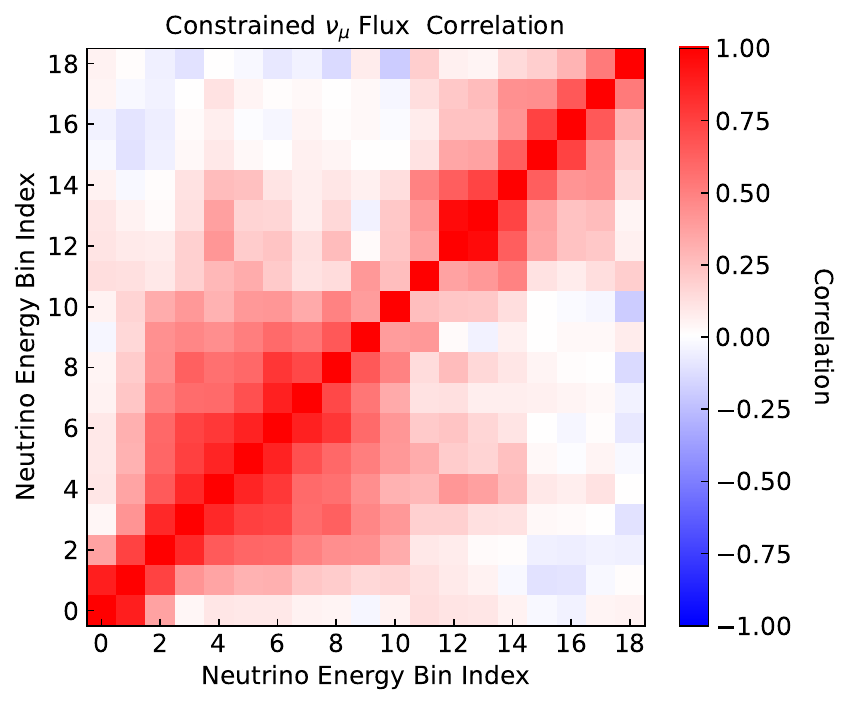}}
    \caption{Pre-constraint (a) covariance and (c) correlation matrix. Post-constraint (b) covariance and (d) correlation matrix.}
\label{fig:flux_cov_cor}
\end{figure*}

\subsubsection{Neutrino Flux Template Scale Factors}

The posterior distributions of the neutrino flux template scale factors are stable and well-described as a normal distribution. As can be observed in Fig.~\ref{fig:fit_cov}, the flux template scale factors correlate strongly with the form factor coefficients, with the sign of the correlation determined by the sign of the coefficient. The lower and higher energy bins of the neutrino templates have larger fit uncertainties due to the limited statistics, and therefore correlate less strongly with the relevant nuisance parameters. 

\subsection{Flux Constraint}
The constraint on the flux prediction is achieved by quantifying the agreement of the neutrino template fit result with the distribution of flux universes. Using the DUNE ND muon neutrino flux prediction in Fig.~\ref{fig:DUNE_numu_flux} and the flux covariance matrix in Fig.~\ref{fig:flux_cov_cor}~ \cite{DUNETDR}, 10,000 model flux universes were generated by randomly drawing from the multivariate normal distribution. The probability that a universe is consistent with the fitted simulation is
\begin{equation} \label{eq:constraint1}
\begin{split}
    P(\vec{N}|\vec{M}) = & \frac{1}{|(2\pi)^{\kappa}\Sigma_N|^{1/2}} \\
    & \times \mathrm{exp} \left[ \frac{1}{2} (\vec{N}-\vec{M})^T \Sigma_N^{-1} (\vec{N}-\vec{M}) \right] ,
\end{split}
\end{equation}
where $\Sigma_N$ is the covariance matrix for the $\CCCoh$ template normalization parameters, $\vec{N}$ is the fitted values of the template normalization parameters, $\vec{M}$ is the relative normalizations of the thrown flux universe to the nominal flux universe, and $\kappa$ is the number of templates. With the calculated probabilities for the $N_k$ universes, the constrained flux covariance matrix elements are calculated as 
\begin{equation} \label{eq:constraint2}
    \Xi_{ij} = \frac{1}{N_k} \Sigma_k [P(\vec{N}|\vec{M})_k (M_{ik} - \bar{M_i})(M_{jk} - \bar{M_j})] ,
\end{equation}
where $\bar{M}_i = \frac{1}{N_k} [\Sigma_k P(\vec{N}|\vec{M})_k M_{ik}]$ is the weighted average in the $i$th bin.

\begin{figure*}[]
    \centering
    \subfloat[\label{sfig:pre_cov_shape}]{
    \includegraphics[width=0.49\linewidth]{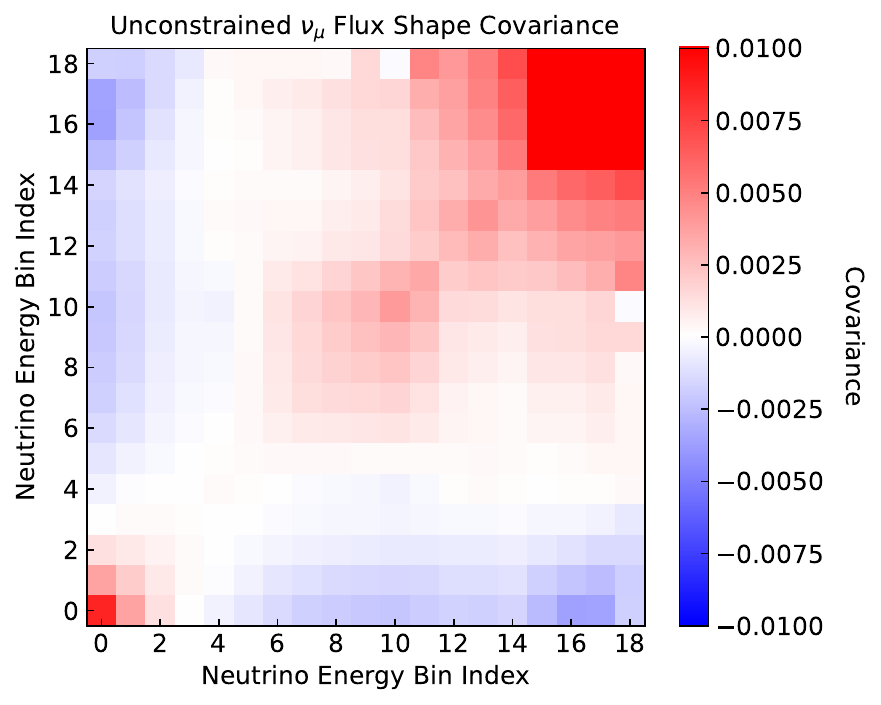}}
    \subfloat[\label{sfig:pre_cov_shape}]{
    \includegraphics[width=0.49\linewidth]{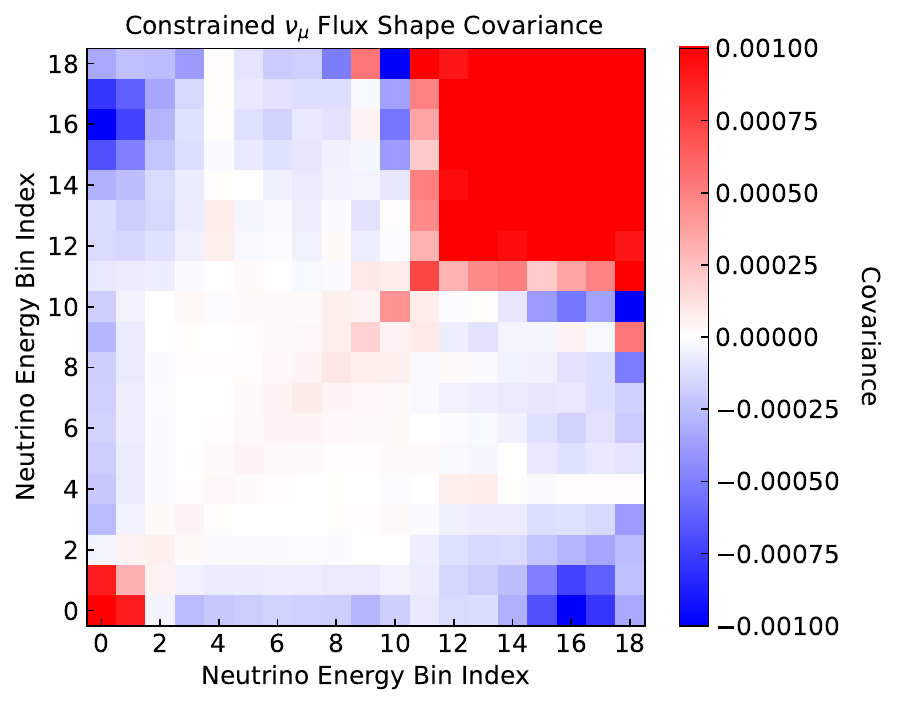}}
    \\
    \subfloat[\label{sfig:pre_cov_shape}]{
    \includegraphics[width=0.48\linewidth]{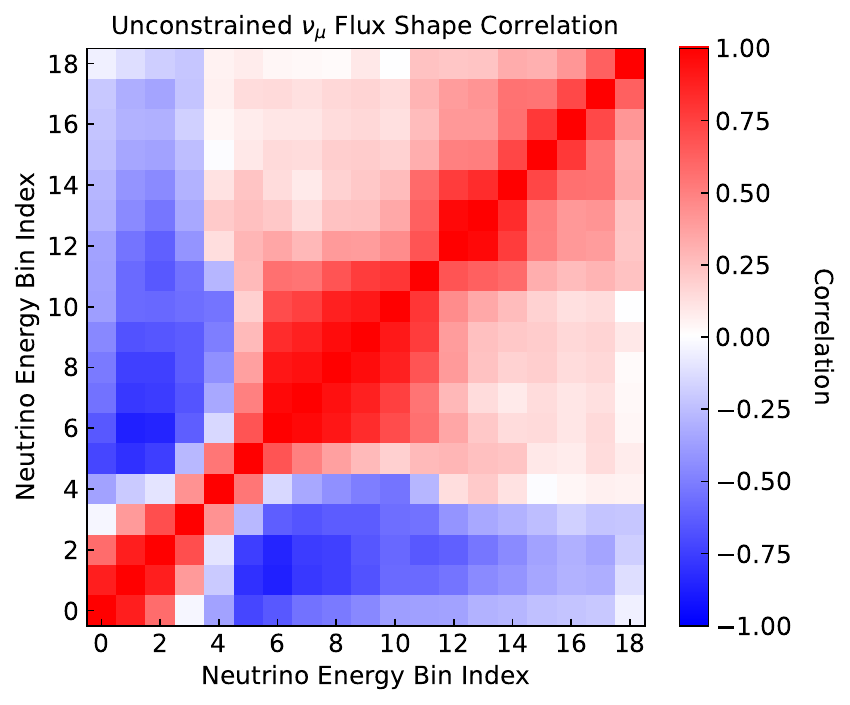}}
    \subfloat[\label{sfig:pre_cov_shape}]{
    \includegraphics[width=0.48\linewidth]{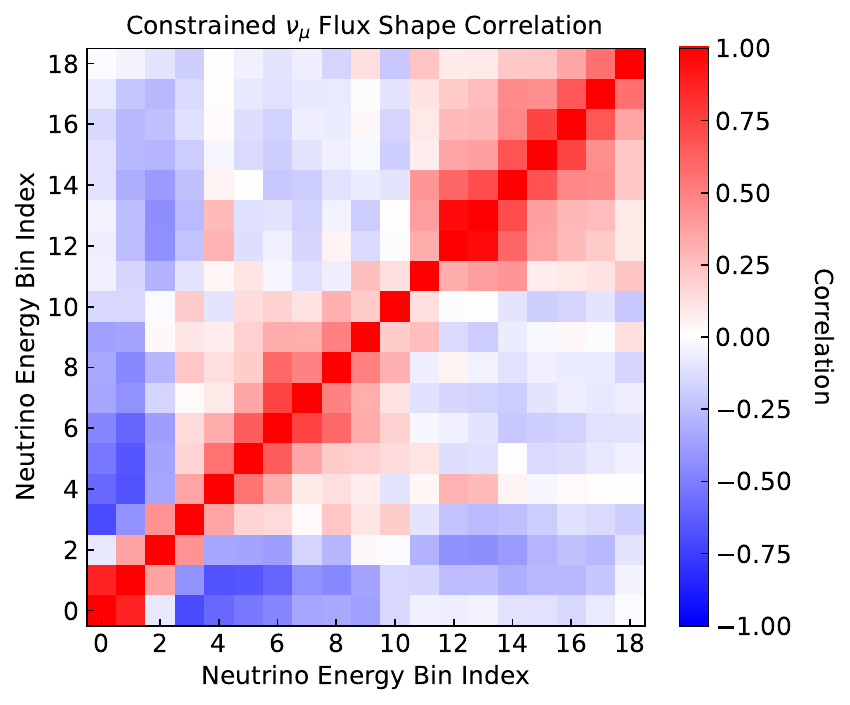}}
    \caption{Pre-constraint shape-only (a) covariance and (c) correlation matrix. Post-constraint shape-only (b) covariance and (d) correlation matrix.}
\label{fig:flux_cov_cor_shape}
\end{figure*}

Figure~\ref{fig:flux_cov_cor} shows the flux covariance matrix and correlation matrix with and without the constraint. The covariance values significantly decrease for most bins, with the exception of the bins at very low and high energy where the statistics are very limited. The correlation also decreases across most bins, especially for bins that are far off-diagonal. While the lower energy bins remain mostly positive correlated with each other, some higher energy bins become anti-correlated after the constraint. The corresponding matrices for only the shape of the distribution is calculated by normalizing the integrated flux for the generated universes and are shown in Fig.~\ref{fig:flux_cov_cor_shape}. We observe the constraint for the shape-only covariance and correlation to be weaker than for the normalization, but still significant.

\begin{figure*}[]
    \centering
    \subfloat[\label{sfig:extpi_unc}]{
    \includegraphics[width=0.49\linewidth]{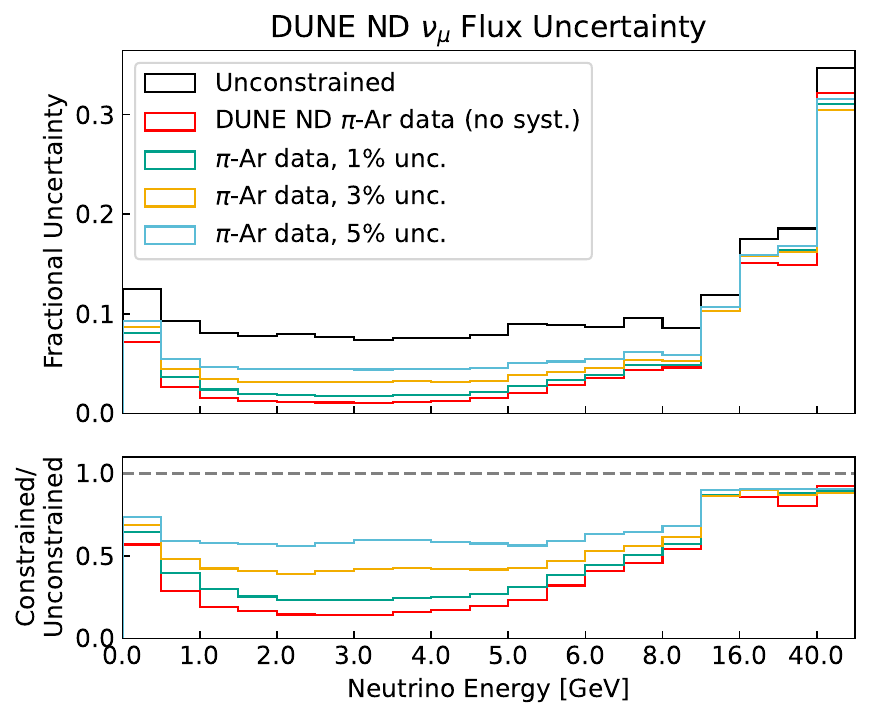}}
    \subfloat[\label{sfig:extpi_unc_shape}]{
    \includegraphics[width=0.49\linewidth]{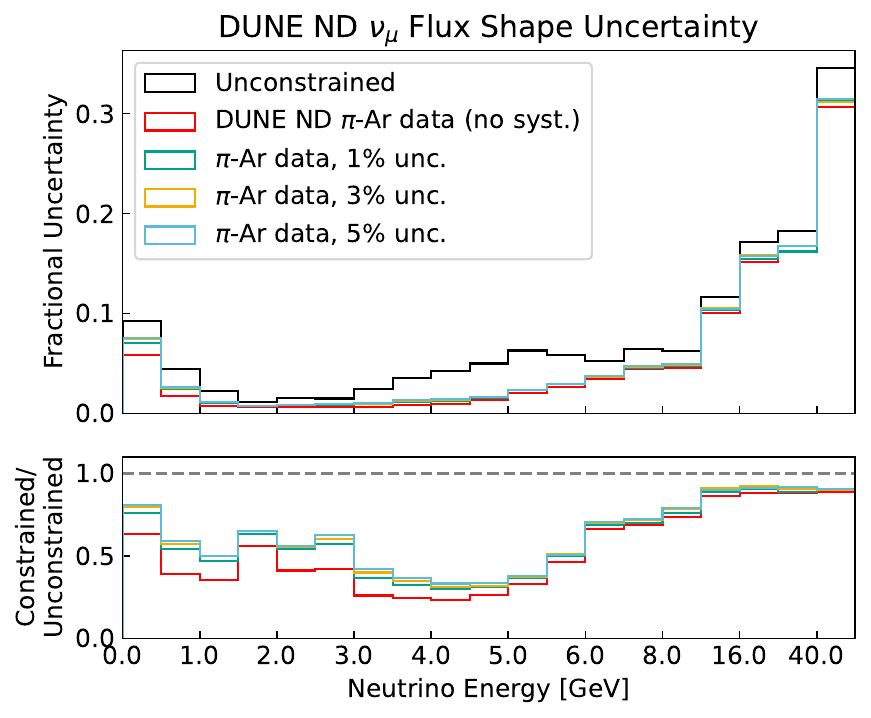}}
    \caption{Unconstrained and constrained flux (a) full uncertainty and (b) shape-only uncertainty for fitting the (stat.~uncertainty only) DUNE ND pion scattering data (the nominal scenario) and assuming a $\pi-$Ar elastic cross section  measurement with a net 1\%, 3\%, and 5\% uncertainty. The latter three scenarios are used solely to parameterize the impact of an assumed overall uncertainty on the pi-Ar elastic cross section, independent of how that uncertainty is obtained.}
    \label{fig:fit_compare_extpi}
\end{figure*}

\begin{figure*}[]
    \centering
    \subfloat[\label{sfig:extpi_unc}]{
    \includegraphics[width=0.49\linewidth]{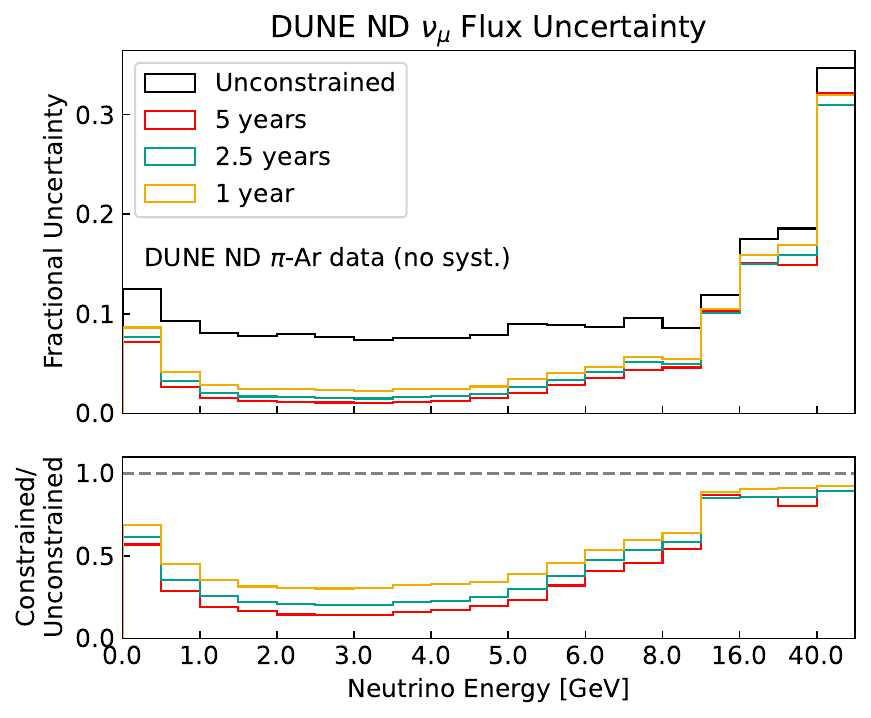}}
    \subfloat[\label{sfig:extpi_unc}]{
    \includegraphics[width=0.49\linewidth]{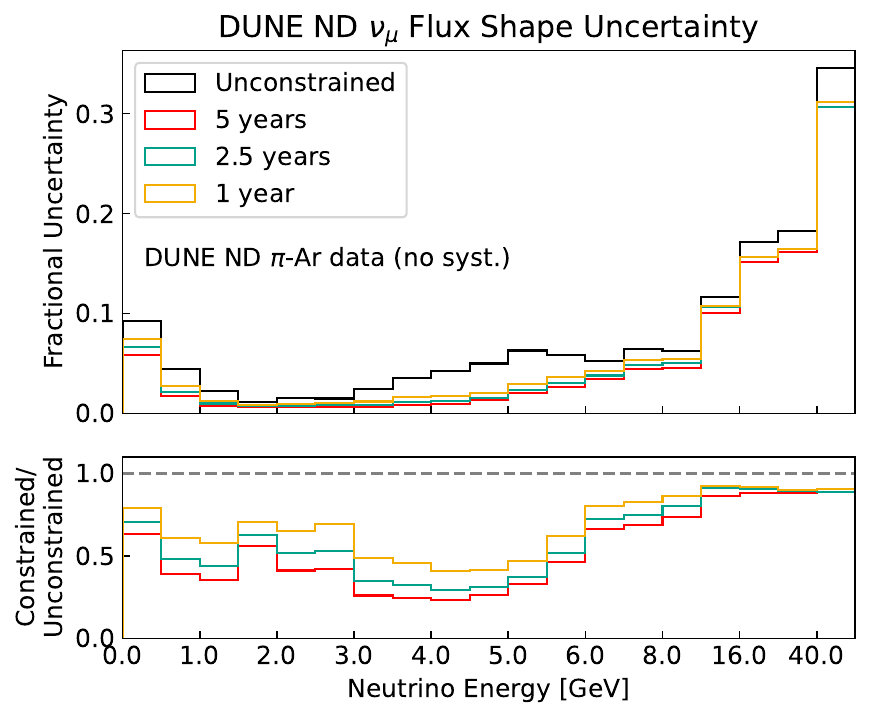}}
    \caption{Unconstrained and constrained flux (a) full uncertainty and (b) shape-only uncertainty for 5, 2.5, and 1 year of running, assuming exposure of 1.1$\times$10$^{21}$ POT per year.}
\label{fig:fit-stat}
\end{figure*}

\begin{figure*}[]
    \centering
    \subfloat[\label{sfig:FFvar_kmax8_FF}]{
    \includegraphics[width=0.49\linewidth]{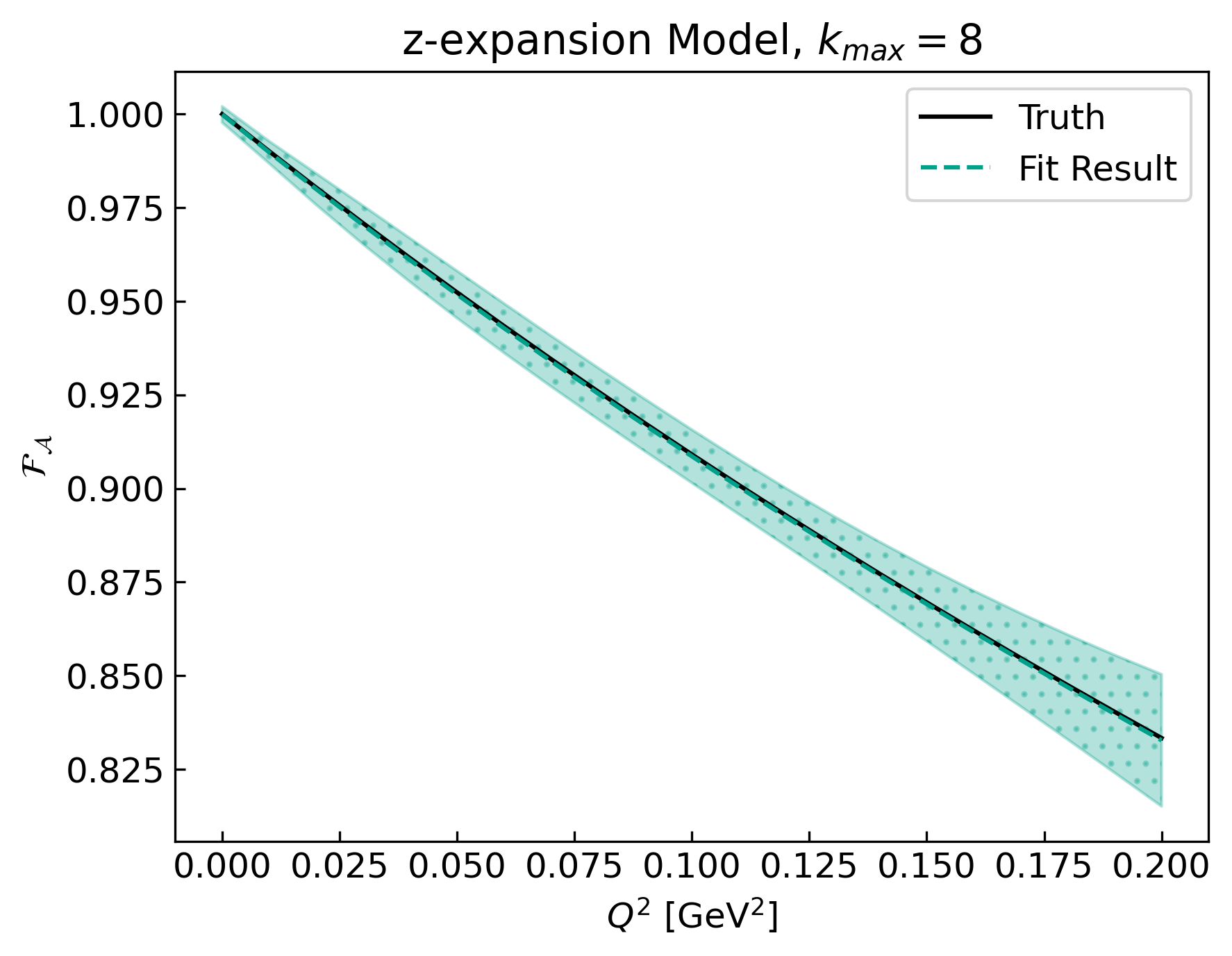}}
    \subfloat[\label{sfig:FFvar_dipole_FF}]{
    \includegraphics[width=0.49\linewidth]{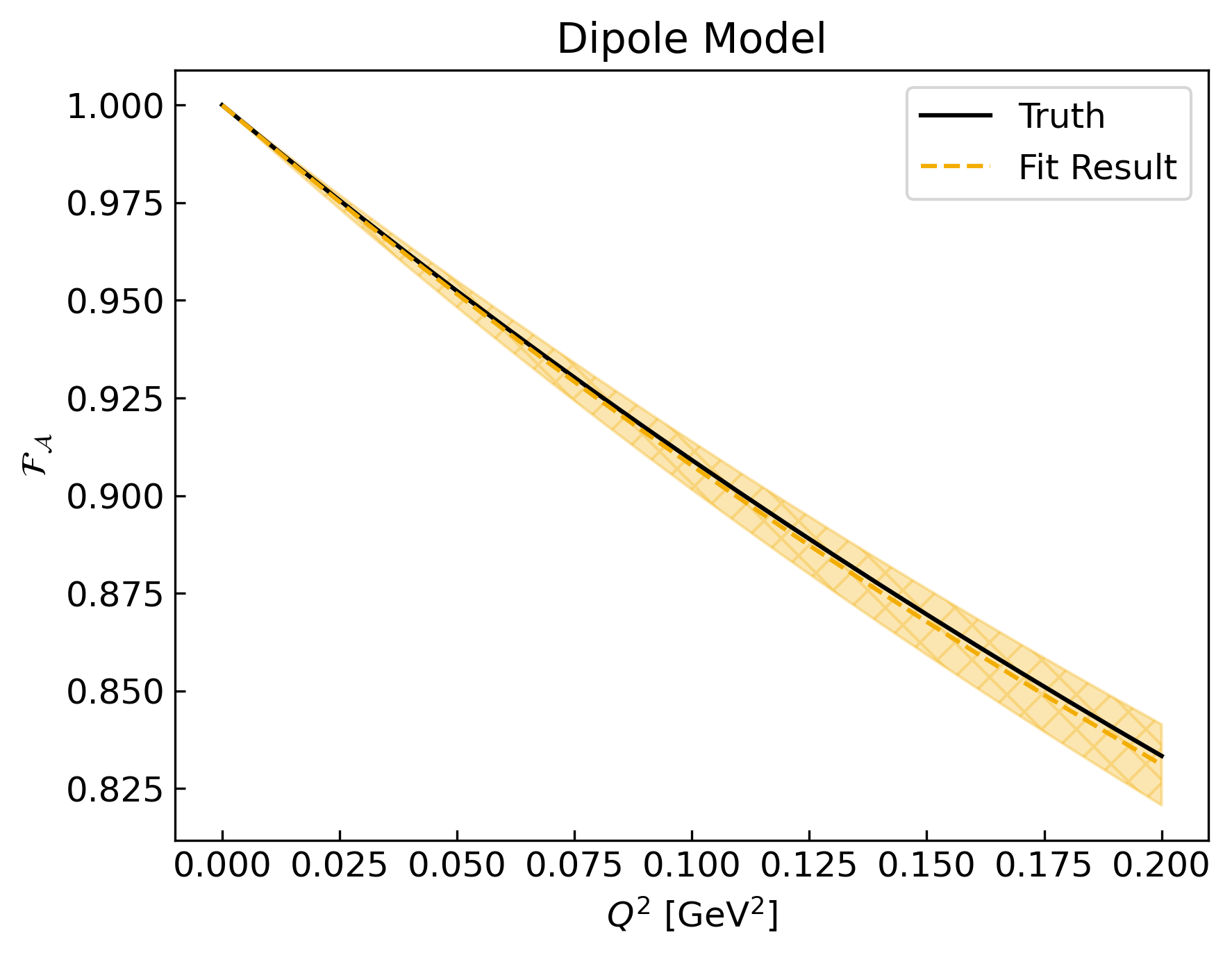}}
    \\
    \subfloat[\label{sfig:FFvar_unc}]{
    \includegraphics[width=0.49\linewidth]{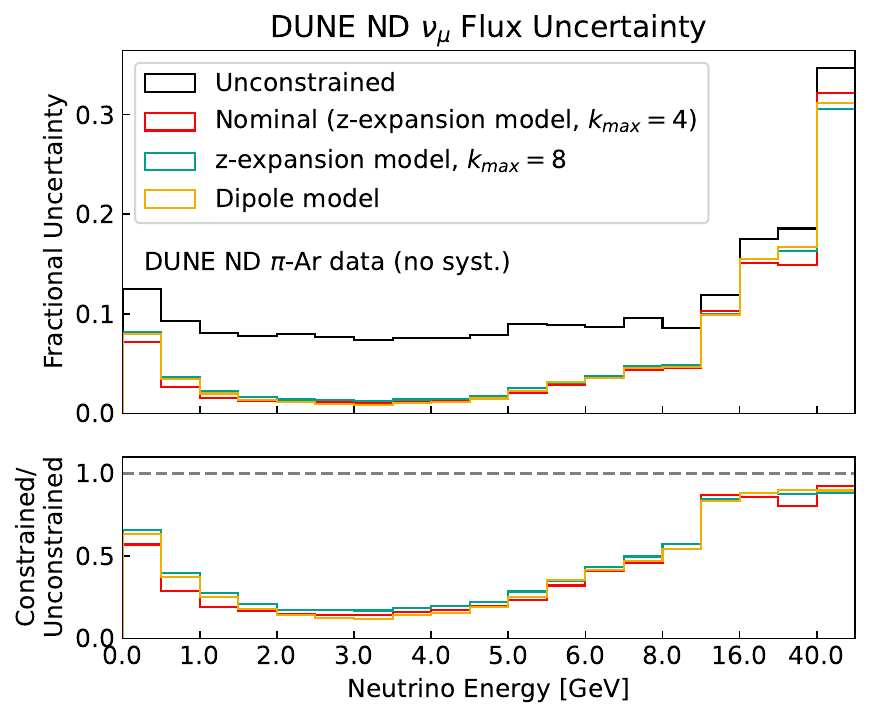}}
    \subfloat[\label{sfig:FFvar_unc_shapeonly}]{
    \includegraphics[width=0.49\linewidth]{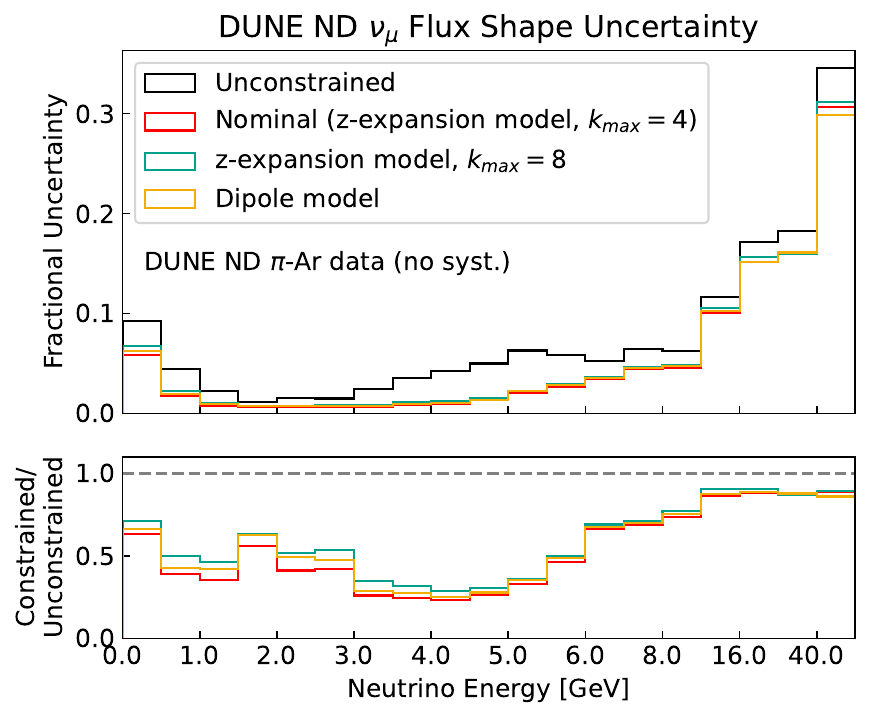}}
    \caption{Form factor fit result using (a) the z-expansion model with 8 coefficients and (b) the dipole model. Unconstrained and constrained flux (c) full uncertainty and (d) shape-only uncertainty using $z$-expansion model with 4 coefficients (the nominal model), $z$-expansion model with 8 coefficients, and dipole model.}
\label{fig:fit_compare_FF}
\end{figure*}

The flux fractional uncertainties are obtained from the diagonal elements of the covariance matrices. Figure~\ref{fig:fit_compare_extpi}. shows the result when we fit the $\pi$-Ar cross section at DUNE ND, along with the results when it is set to an external prior. The constraint is strongest for bins around the oscillation maxima, and weaker for lower and higher energy bins that have very limited statistics. 

For the external $\pi$-Ar scenarios, we test three prior scenarios: 1\%, 3\%, and 5\%. For each case, the $\pi$-Ar template binning (Table~\ref{table:pibins}) is replaced with a single scale factor. This factor is intended to parameterize the overall uncertainty on the pion–argon elastic scattering cross section rather than its detailed kinematic dependence. This simplification is motivated by the fact that, even in the more realistic DUNE ND only scenario, the $\pi$-Ar scattering rate uncertainty is almost completely correlated between different neutrino energy bins (see Sec.~\ref{subsec:PiArConstraint}, Fig.~\ref{fig:pixsec_nu_fit}). Furthermore, we tested more complicated versions of the prior on the $\pi$-Ar cross section scale factor, and found that the resulting flux constraint was mostly sensitive to the net normalization uncertainty.


In the DUNE ND $\pi$-Ar scenario (which neglects systematic uncertainties on the elastic cross section), the uncertainty on the neutrino flux at its peak is constrained to $\sim$1.5\%, a reduction of $\sim \,$6 times relative to the prior uncertainty. In the external $\pi$-Ar data scenario, the uncertainty on the neutrino flux at its peak is constrained to $\sim 2$-$5$\%, for a $\pi$-Ar elastic scattering cross section constraint of $1$-$5$\%. The external $\pi$-Ar scenario can also be used to interpret the impact of any systematic uncertainty in the DUNE ND measurement of the $\pi$-Ar elastic cross section on the extraction of the neutrino flux. That is, if systematic uncertainties limit the DUNE ND $\pi$-Ar elastic cross section measurement to an uncertainty of (e.g.) 3\%, that is equivalent to an external measurement with an uncertainty of 3\% (up to correlations, which are not included in this study). 
It should be noted that these results should not be used to directly compare the projected performance of a flux constraint from a DUNE-ND $\pi-$Ar elastic cross section measurement to an external measurement. In particular, projecting the relative size of the (likely dominant) systematic uncertainties from the two measurements is beyond the scope of this work.


Figure~\ref{fig:fit-stat} shows the constraint for 5, 2.5, and 1 years of running for the DUNE ND $\pi$-Ar cross section scenario. For 5 years of running, the uncertainty is reduced from $\sim 8$\% to $\sim 1.5$\% at its peak, and the shape-only uncertainty is halved. While the full 5-year running period provides the most powerful flux constraint, even a single year of data yields a significant constraint. The variance in runtime can also be used to understand the impact of a non-idealized reconstruction efficiency less than 100\%, as we have assumed in this analysis.

We test the impact of the axial-vector form factor model choice on the flux constraint by repeating the fit procedure using two variations of the model: the dipole model with the axial mass as the only model parameter (as used by GENIE) and the $z$-expansion model with 8 coefficients. The form factor fit result in Fig.~\ref{fig:fit_compare_FF} shows that the fit using the dipole model results in smaller fit uncertainties, and the fit result using the $z$-expansion model with 8 coefficients is comparable to the nominal case. Figure~\ref{fig:fit_compare_FF} shows that the flux constraint performance is similar across the variations. Although not shown here, the fit was also repeated with different $t_0$ values for the $z$-expansion model: 0 and \SI{-0.2}{GeV\squared} in place of the nominal value of \SI{-0.07}{GeV\squared}, which was optimized specifically for the $Q^2$ distribution of the simulated $\CCCoh$ events. In both cases, no significant effects on the result were found.

\subsection{Bias Tests}

\begin{figure}[]
    \centering
    \subfloat[\label{sfig:Mavar_08}]{
    \includegraphics[width=\linewidth]{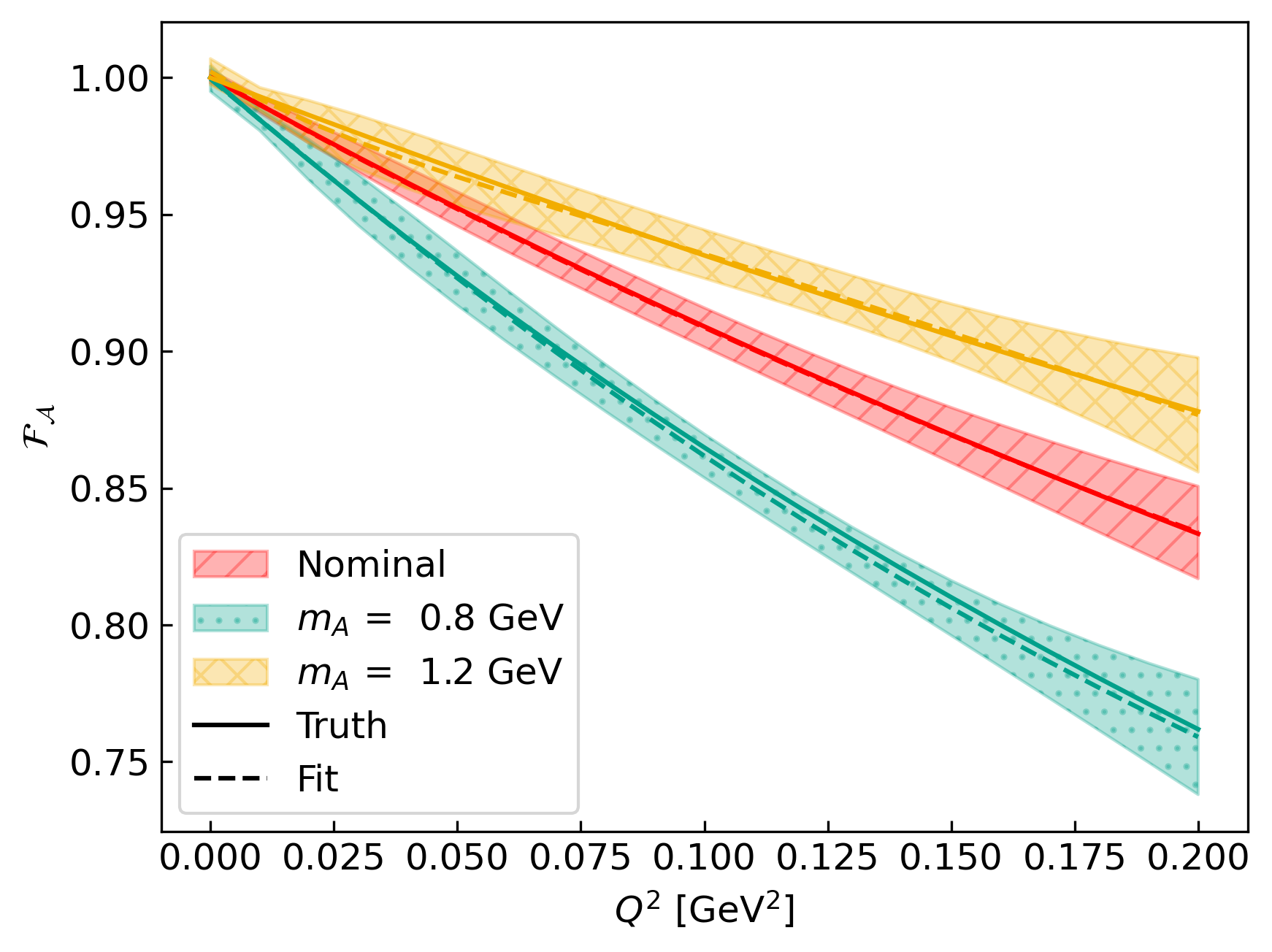}} \\
    \subfloat[\label{sfig:Mavar_bias}]{
    \includegraphics[width=\linewidth]{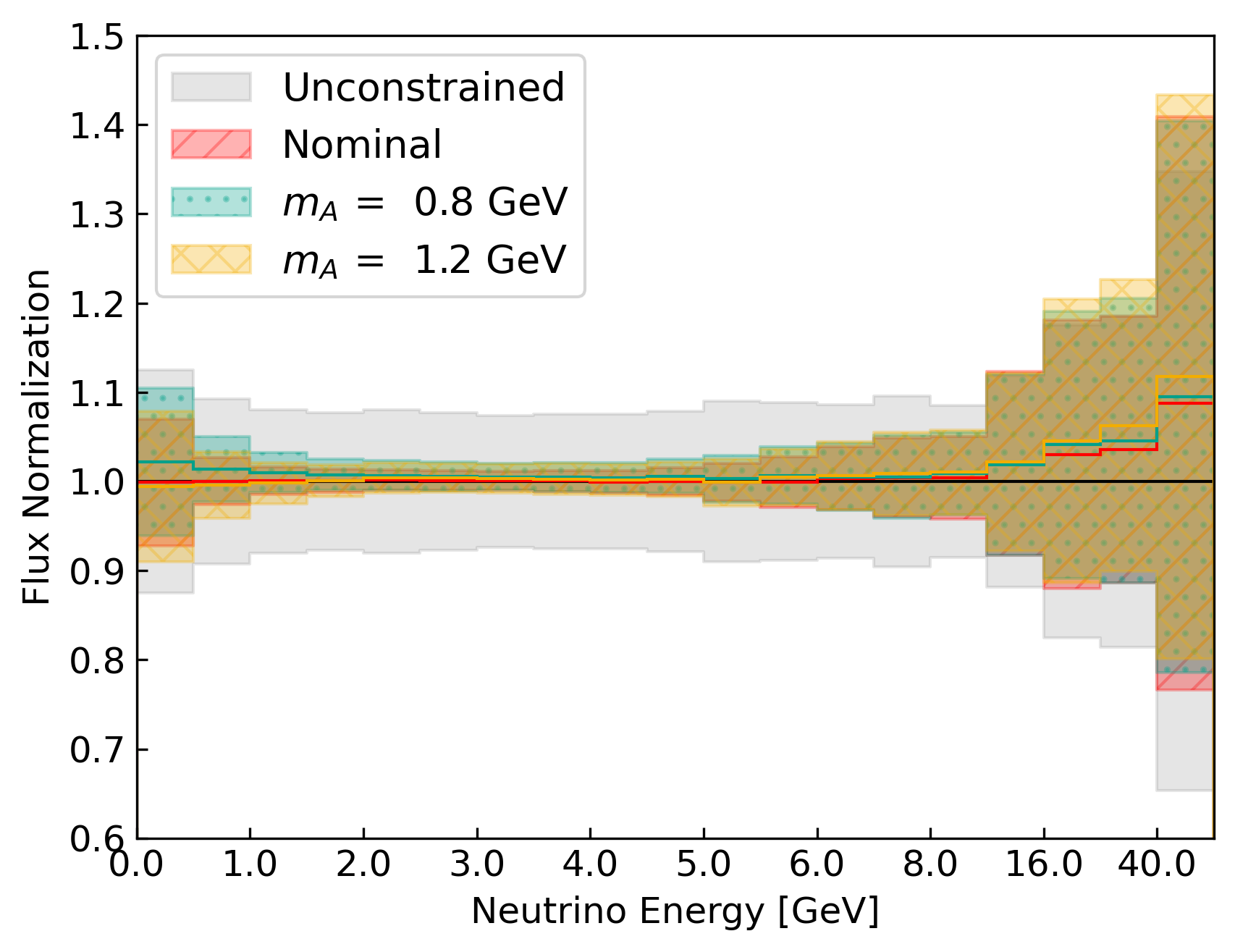}}
    \caption{Fit results for the form factor (a) and flux normalization (b), obtained using fake datasets generated with a dipole form factor model assuming axial mass values of 0.8$\,$GeV and 1.2$\,$GeV. Solid line marks the weighted average, and the colored area spans the bin-by-bin standard deviation.}
\label{fig:fit_compare_datavar_MA}
\end{figure}

\begin{figure}
    \centering
    \includegraphics[width=\linewidth]{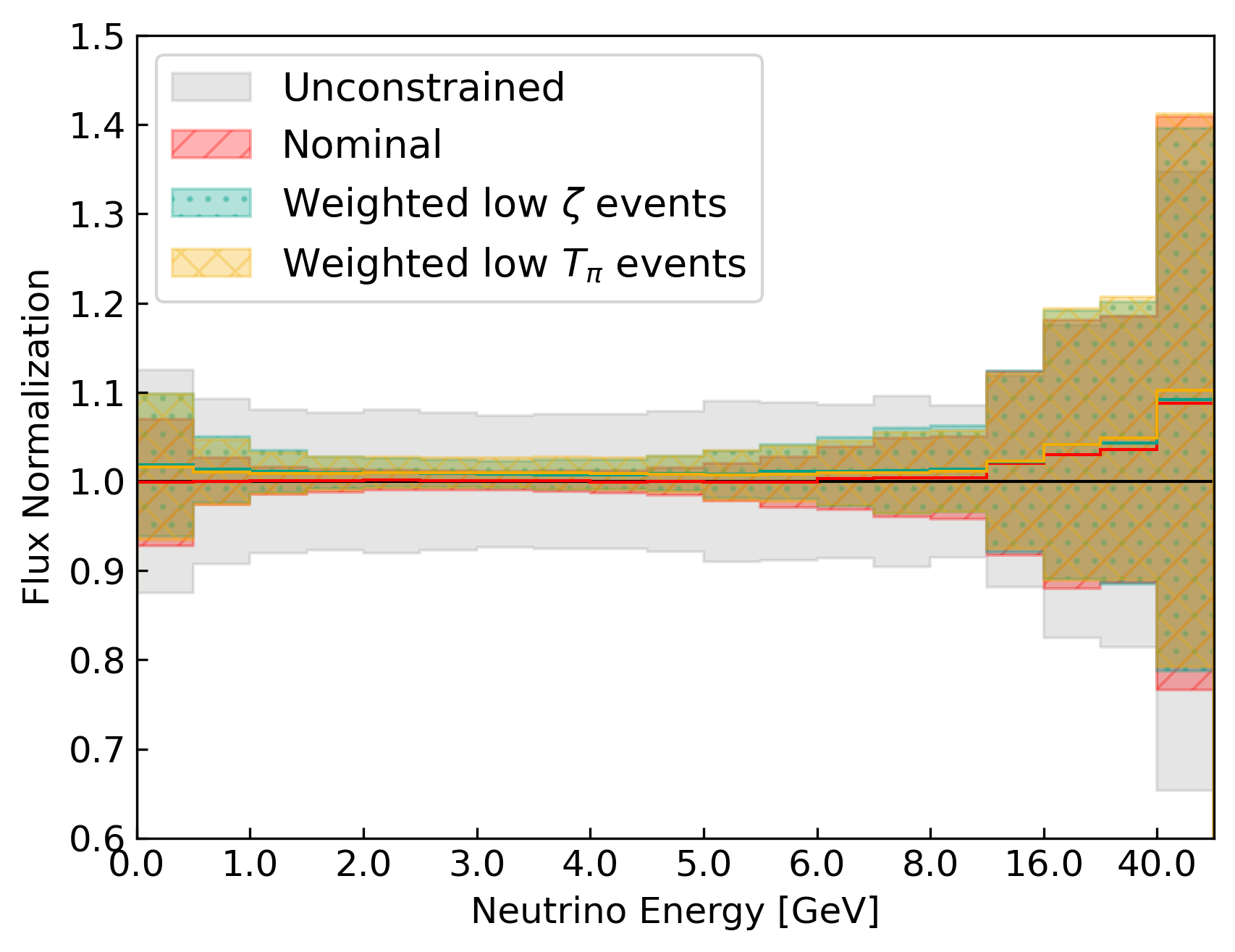}
    \caption{Fit results for the flux normalization, obtained using fake datasets generated by weighting events with $\zeta < 3$ and $E_{\pi}< 0.25\,$GeV by a factor of 0.7. Solid line marks the weighted average, and the colored area spans the bin-by-bin standard deviation.}
\label{fig:fit_compare_datavar_weightevts}
\end{figure}

\begin{figure}
    \centering
    \includegraphics[width=\linewidth]{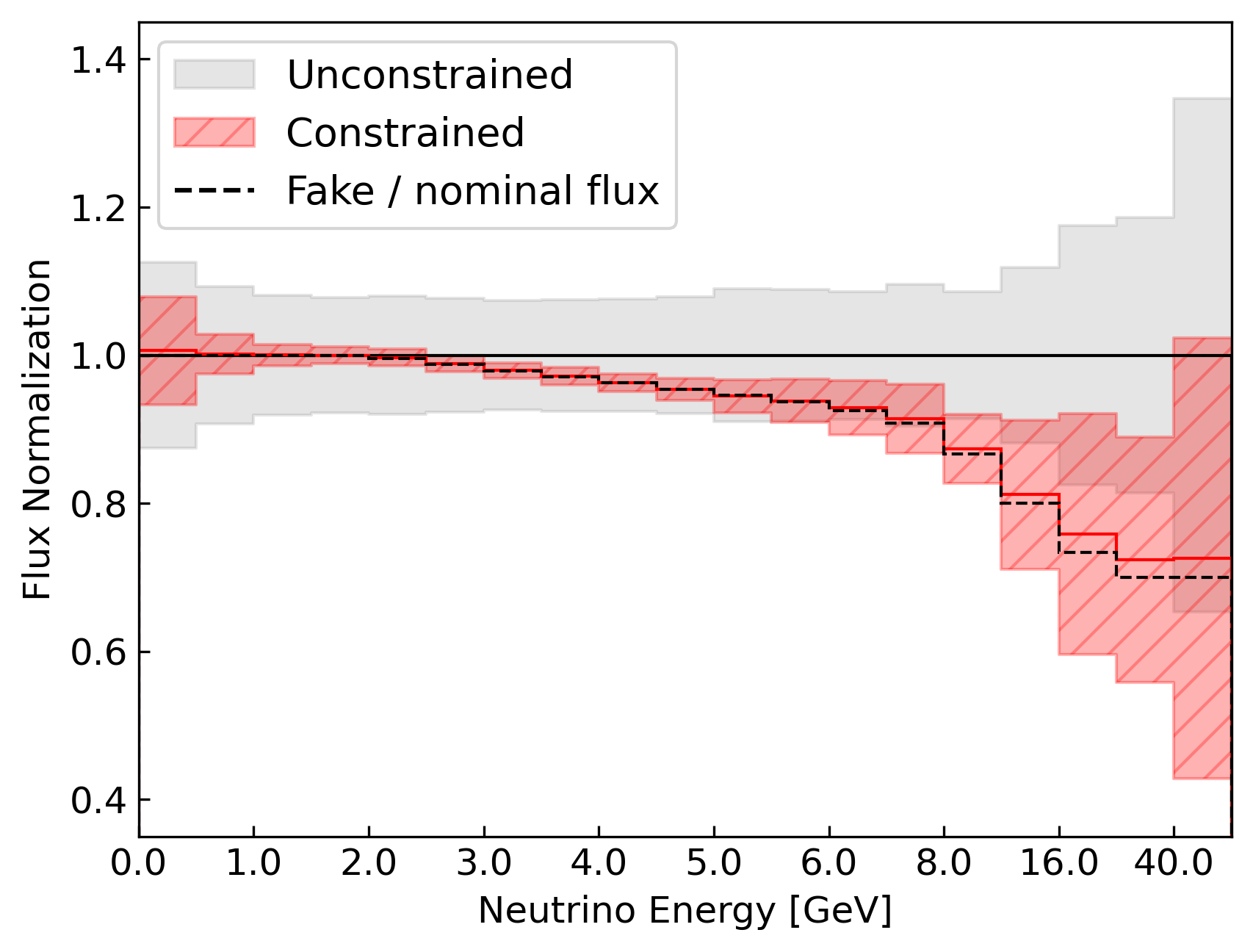}
    \caption{Fit results for the flux normalization, obtained using fake dataset generated using a fake flux, shown as a ration to the nominal flux model with black dotted line. Solid line marks the weighted average, and the colored area spans the bin-by-bin standard deviation.}
    \label{fig:fit_compare_datavar_fakeflux}
\end{figure}

To test the resilience of the proposed method to potential systematical effects that may cause discrepancies between data and simulation, the fit procedure was repeated using fake datasets. First, we repeat the fit using fake datasets generated with different dipole form factor axial mass values: 0.8 GeV and 1.2 GeV. This test accounts for possible variations in the axial contribution to the cross-section relevant at non-zero $Q^2$~\cite{Paschos1, PCACBad}. The test variation is about the same in magnitude as those estimates (e.g., 5\% at $Q^2=$ \SI{0.1}{GeV\squared}). Figure~\ref{fig:fit_compare_datavar_MA} shows the form factor fit result for these cases, as well as the average and the standard distribution of the flux normalizations of the weighted universes. The results demonstrate that although the fit central values deviate slightly from the simulation truth values, the method is able to accommodate these deviations within the fit uncertainties. The flux uncertainty constraint performance is not significantly affected. 

Next, we generate fake datasets by weighting by a factor of 0.7 the events in the kinematic phase space where the Adler relation is not reliable: $\zeta < 3$ and $E_{\pi}< 0.25\,$GeV. Figure \ref{fig:fit_compare_datavar_weightevts} show that the flux constraint performance remains unaffected in these cases as well. This test demonstrates the robustness of the fit against the systematic uncertainties in the phase space that are more strongly influenced by the transverse currents and the effects of the sign difference.

Lastly, we test the scenario where the true flux differs substantially from the prior model. We generate a fake dataset using a flux model scaled from the nominal model, with no modification below \SI{2}{GeV}, a linear reduction increasing from 0\% at \SI{2}{GeV} to 30\% at \SI{20}{GeV}, and a constant 30\% for above \SI{20}{GeV}. This choice is motivated by the DUNE flux produced by decreasing the target density by 30\%, reported in Ref.~\cite{DUNEConstraint}. Figure \ref{fig:fit_compare_datavar_fakeflux} demonstrates that the method successfully fits to this flux model, remaining well within the constrained uncertainties. This result further demonstrates that our approach can directly measure the flux, in addition to constraining its uncertainties.

\section{Conclusion}
\label{sec:conclusion}
The upcoming DUNE program has the ambitious goal of observing charge-parity violation in long-baseline neutrino oscillations. The design of the experiment powerfully constrains the impact of significant systematic uncertainties on the neutrino cross section and flux inherent to neutrino experiments. Still, techniques which can directly constrain the neutrino flux in-situ will enhance the sensitivity of DUNE. 

In this work, we have demonstrated that the unique properties of the $\CCCoh$ neutrino-nucleus interaction, as well as the capabilities of the DUNE ND, could enable such a flux constraint. Furthermore, we have developed a method that provides a reliable flux constraint while accounting for the systematic effects on the $\CCCoh$ cross section from the axial-vector form factor and the $\pi$-Ar scattering cross section. We find that the method has the statistical power to reduce the fractional uncertainty on the muon neutrino flux from about 10\% to 1.5-5\% near the oscillation maxima energy, depending on how well the $\pi$-Ar elastic cross section can be measured. An overall elastic cross section measurement uncertainty of 1-5\% (whether due to systematic uncertainties in DUNE ND, or the resulting uncertainty from an external measurement), would constrain the flux peak fractional uncertainty to $\sim 2$-$5$\%. These results neglect any uncertainty associated with the DUNE ND detector performance. They also  do not include a number of other experimental considerations, such as the wrong-sign flux component, which would necessarily be a part of a real measurement.

This study demonstrates the possible utility of a $\CCCoh$ flux-constraint, but there are still outstanding questions that need to be resolved before it can be realized in practice. In particular, more study is needed to elucidate the range of validity of the Adler relation, and if there are any necessary corrections or uncertainties at the DUNE neutrino flux energy. Measurements of the $\CCCoh$ process on argon in ongoing experiments, such as the SBN program, would aid this effort. In addition, measurements of the elastic $\pi$-Ar cross section, such as at ProtoDUNE or LArIAT, would further demonstrate the feasibility of this method and provide necessary input to any study of the $\CCCoh$ process. In addition to its utility as a flux constraint, studying the $\CCCoh$ process can shed light on the fundamental nature of the axial current in neutrino interactions, as expressed by the PCAC theorem and the Adler relation.

\acknowledgments
We would like to acknowledge the helpful discussions with Ryan Plestid and Alexis Nikolakopoulos in preparation of this manuscript. We also thank Jonathan Rositas for contributions to the form factor studies during the early stages of this project. We gratefully acknowledge the valuable feedback on this manuscript from the members of the DUNE collaboration, particularly Richard Diurba, Laura Fields, Yoann Kermaidic, Yinrui Liu, Kendall Mahn, Christopher Marshall,
Sungbin Oh, Roberto Petti, Aaron Higuera Pichardo, Leigh Whitehead, Lee Hagaman, and Jeremy Wolcott. This manuscript has been authored by Fermi Forward Discovery Group, LLC under Contract No. 89243024CSC000002 with the U.S. Department of Energy, Office of Science, Office of High Energy Physics. This material is based upon work supported by the National Science Foundation under Grant No. PHY-2209601. This work was made possible through the support of the Enrico Fermi Fellowships led by the Center for Spacetime and the Quantum, and supported by Grant ID \#63132 from the John Templeton Foundation. The opinions expressed in this publication are those of the author(s) and do not necessarily reflect the views of the John Templeton Foundation or those of the Center for Spacetime and the Quantum.

\bibliography{cohpi}

\end{document}